%% file: paper2-arxiv.tex
\def\cleancopy{1} \def\arxivcopy{1}
\documentclass[e-only,10pt,reqno]{ofj}

\usepackage{tensorNotation}
\usepackage{finiteVolumeNotation}

\usepackage{setspace}
\usepackage{lineno}
\usepackage{color}
\usepackage[dvipsnames,svgnames,x11names]{xcolor}
\usepackage[hang,flushmargin]{footmisc}
\usepackage{orcidlink}
\usepackage{cite}
\usepackage{xspace} 

\ifdefined\cleancopy\else\ifdefined\review\else
\usepackage{showlabels}

\fi\fi

\usepackage[colorlinks=true,linkcolor=black,anchorcolor=black,citecolor=black,filecolor=black,menucolor=black,runcolor=black,urlcolor=black]{hyperref}

\newcommand{\authorcontributions}[1]{%
\vspace{6pt}\noindent{\fontsize{9}{11.2}\selectfont\textbf{Author Contributions:} {#1}\par}}
\newcommand{\dataavailability}[1]{%
\vspace{6pt}\noindent{\fontsize{9}{11.2}\selectfont\textbf{Data Availability:} {#1}\par}}

\let\paragraph\undefined

\renewcommand{\datename}{$^{*}$ Corresponding author}
\copyrightinfo{2026}{The Authors. This work is an open access article under the {\color{blue}{\href{https://creativecommons.org/licenses/by-sa/4.0/}{CC BY-SA 4.0}}} license}
\newcommand{\OF}[0]{OpenFOAM\textsuperscript{\textregistered}\xspace}

\usepackage{amssymb} 
\usepackage{float}   
\usepackage{placeins} 
\newcommand{\coderoot}{../fluxinator}

\newcommand{\ptwo}{figures}

\IfFileExists{ulem.sty}
  {\usepackage[normalem]{ulem}}
  {\providecommand{\sout}[1]{{\color{red}##1}}}
\ifdefined\cleancopy
\newcommand{\del}[1]{\ignorespaces}

\newcommand{\movednote}[1]{\ignorespaces}
\newenvironment{addblock}{}{}
\else
\newcommand{\del}[1]{\sout{#1}}

\newcommand{\movednote}[1]{{\color{red}\itshape [#1]}}

\fi

\ifdefined\review
\DeclareRobustCommand{\datasource}[1]{\nolinkurl{#1}}
\else\ifdefined\arxivcopy
\DeclareRobustCommand{\datasource}[1]{\href{https://github.com/isoAdvector/fluxinator/tree/main/#1}{\nolinkurl{#1}}}
\else
\DeclareRobustCommand{\datasource}[1]{\href{run:\coderoot/#1}{\nolinkurl{#1}}}
\fi\fi
\newcommand{\datanote}[1]{\textit{Data:} \datasource{#1}}
\newcommand{\datafn}[1]{\footnote{\datanote{#1}.}}
\newcommand{\figdir}[1]{\textit{Files:} \datasource{papers/part2/#1}}
\ifdefined\review
\newcommand{\selfcite}[1]{\cite{#1Anon}}
\else
\newcommand{\selfcite}[1]{\cite{#1}}
\fi

\newcommand{\parhead}[1]{\par\addvspace{\medskipamount}\noindent{\bfseries #1}\enspace\ignorespaces}

\OpenFOAMversions{\OF v2512}

\ifdefined\review
\Repository{}
\else
\Repository{https://github.com/isoAdvector/fluxinator}
\fi

\DOI{TBD} 

\ifdefined\arxivcopy
\usepackage{etoolbox}
\makeatletter
\let\publname\undefined
\patchcmd{\@maketitle}{\item[DOI] \href{https://doi.org/\theDOI}{\@DOI}\\}{}{}{\PackageError{paper2}{arXiv title patch failed}{}}
\makeatother
\fi

\ifdefined\review
\fi

\begin{document}


\title[Consistent interfacial flow numerics -- Part II]{Consistent interfacial flow numerics -- Part II: The momentum convection term}

\ifdefined\review
  \author{}
  \address{}
  \email{}
\else
\author[J. Roenby]{Johan Roenby\,\orcidlink{0000-0002-7254-0546}$^{1,2,*}$}
\address{$^1$STROMNING ApS, Luftmarinegade 62, 1432 K{\o}benhavn K, Denmark}
\email{johan@stromning.com}
\address{$^2$Department of Science and Environment, Roskilde University, Universitetsvej 1,
  4000 Roskilde, Denmark}
\fi


\input{paper2/abstract.tex}

\ifdefined\review\date{}\else\date{\today}\fi

\dedicatory{}

\maketitle

\ifdefined\arxivcopy\else
  \linenumbers
\fi

\input{paper2/introduction}
\input{paper2/method}
\input{paper2/timeLag}
\input{paper2/simpleFlows}
\input{paper2/interface}
\input{paper2/wave}
\input{paper2/finalSolver}

\input{paper2/conclusion}


\ifdefined\review
\else
  \input{parts/acknowledgements.tex}

  \authorcontributions{
  J.R.\ carried out all parts of the work: conceptualisation, methodology, software, validation,
  formal analysis, investigation, data curation, writing (original draft, review and editing),
  visualisation and funding acquisition. The author has read and agreed to the published version
  of the manuscript.
  }

  \ifdefined\arxivcopy
  \dataavailability{
  The solver, the case set-ups and the post-processing scripts are available at
  \url{https://github.com/isoAdvector/fluxinator}. The repository will be archived at Zenodo
  on publication.
  }
  \else
  \dataavailability{
  The solver, the case set-ups and the post-processing scripts are archived at Zenodo
  \cite{RoenbyData2026}. The code is
  developed further at \url{https://github.com/isoAdvector/fluxinator}.
  }
  \fi
\fi

\ifdefined\review
  \dataavailability{
  The solver, the case set-ups and the post-processing scripts are in the supplementary file,
  whose \texttt{README\_REVIEW.md} explains how to build and run them. The paths in the captions
  (\emph{Files}) and footnotes (\emph{Data}) refer to it. The repository will be archived at Zenodo
  on publication.
  }
\fi

\input{parts/aiDeclaration.tex}


\bibliographystyle{IEEEtran}

\bibliography{Bibliography}

\end{document}

%% file: paper2/abstract.tex
\begin{abstract}
The two-phase Volume-of-Fluid solvers of \OF overpredict the horizontal water velocity near
the crest of a water wave, and the error grows as the wave propagates. Part~I of this series
removed the spurious air velocities that the gravity term causes at an interface at rest and reduced them in waves, but left this error in the water column unchanged.
Here we show that it is caused by the momentum convection term. In the \OF solvers the mass
flux of the convection term is computed in the interface advection step from the velocity at the
beginning of the time step. The convected velocity, however, is taken at the end of the step, because
the convection term is discretised implicitly. In the equation for the horizontal velocity, the vertical velocity that transports
it is therefore one time step older than the horizontal velocity. Since the two are 90 degrees out of phase in a progressive wave, this time lag produces an upward flux of horizontal momentum, proportional to the time
step, which accumulates in a spurious current in the water near the surface. The mass flux also carries
the density of the start of the step, so the same mismatch makes the response to a force depend
on the density at the wrong time level. Evaluating the convected velocity at the start of the
step (explicit convection) removes the lag and makes the response consistent with the gravity
force. Explicit convection, however, requires the convected face velocity at the interface to be taken
from the upwind cell. Only then is the new velocity of a cell a weighted average of old
velocities, whatever the density ratio. Other convected face velocities amplify velocity differences in
cells that lose most of their water in a step. We demonstrate both requirements
with prescribed flows, and confirm with the full solver that explicit convection reproduces a uniformly
accelerated water disc to the tolerance of the pressure solver. For a progressive wave, explicit convection with upwind convected face velocities at the
interface reduces the overprediction of the crest velocity from 26--31\,\% to within 1\,\% and
the current in the water near the surface to zero. The resulting solver, \texttt{fluxinator}, is meant for maximum
Courant numbers of 0.1 and below. There it keeps the height of the tested wave to within 1.5\,\% over 15 periods
on meshes with six or more cells per wave height, whereas at larger Courant numbers the wave grows.
Because the spurious velocities that limit the time step are removed, \texttt{fluxinator} needs 34--42\,\% less
wall-clock time than \texttt{interFoam} at a maximum Courant number of 0.1 for the tested progressive wave case.
\end{abstract}

%% file: paper2/introduction.tex
\section{Introduction}
\label{sec:introduction}

Two-phase Volume-of-Fluid (VoF) solvers that compute both water and air are widely used for
water waves. Those of \OF are known to overpredict the horizontal water velocity near the wave
crest. Wroniszewski et al.\ \cite{Wroniszewski2014} found a strong overprediction of the
velocity under the crest of a solitary wave with \texttt{interFoam} and suggested that it may be due to
the interface being smeared over several cells, which makes the flow resemble a continuous
stratification. Larsen et al.\ \cite{Larsen2019}
reported the problem for progressive waves with \texttt{interFoam}, and Qwist and Christensen
\cite{Qwist2023} found an overprediction of 20--30\,\% for \texttt{interIsoFoam}\footnote{\texttt{interIsoFoam} is essentially \texttt{interFoam} with the geometric
isoAdvector method \cite{Roenby2016} instead of the algebraic MULES scheme for
the interface advection.} that did not decrease
with mesh refinement. Qwist and Christensen removed the error with a method that does not
compute the air phase, and noted that a ghost fluid method (GFM) \cite{Fedkiw1999,Kang2000,Vukcevic2017} that retains the air overpredicts
the crest velocity as well. Among the possible causes, they suggested that the sharp
interface of \texttt{interIsoFoam} is inconsistent with a velocity that is continuous across it.

\begin{figure}[tbp]
\centering
\includegraphics[width=\linewidth]{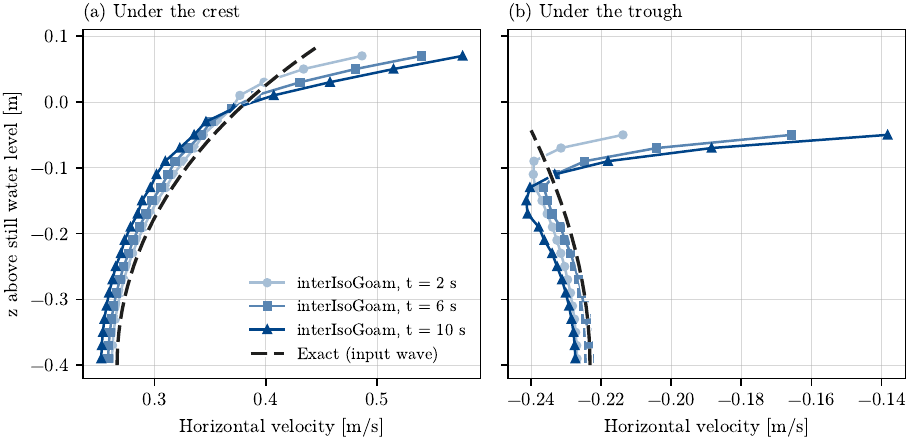}
\caption{The problem addressed in this paper: progressive wave of Section~\ref{sec:wave}
computed with \texttt{interIsoGoam} and its hardcoded implicit momentum convection ($190\times40$ mesh,
$\Delta t = 0.005$\,s). Horizontal velocity in the water ($\alpha \ge 0.5$) in the column nearest (a) the
crest and (b) the trough, after one, three and five periods. The dashed line is the exact wave of stream
function theory with which the run is initialised (the exact input wave).
\figdir{fig_problem}.}
\label{fig:problem}
\end{figure}

Part~I of this series \selfcite{RoenbyPart1} showed that the gravity term of the \OF solvers
produces spurious velocities near the interface, mainly in the air. Evaluating the gravity force
at the height of the reconstructed interface removed them for a horizontal interface at rest and
reduced them in waves. The overprediction of the crest velocity was nearly
the same with and without that correction. For all three solvers, \texttt{interFoam}, \texttt{interIsoFoam} and
\texttt{interIsoGoam}\footnote{\texttt{interIsoGoam} is the solver of Part~I: \texttt{interIsoFoam} with a gravity term that
uses the centroids of the reconstructed interface segments as the interface height (the
geometric gravity model).}, the horizontal velocity in the top full water cell under the crest exceeded the
exact value by 45--60\,\% after 15 wave periods, with the \texttt{ddtCorr} term of \OF on, as it
is by default. \texttt{interFoam} smears the interface over several cells, while \texttt{interIsoFoam} and \texttt{interIsoGoam}
keep it sharp. Yet in the Part~I wave study, repeated in Section~\ref{sec:final} of the
present paper, \texttt{interFoam} overpredicts the speed under the crest as well, by 18--107\,\% of the
exact crest velocity, and in most runs by no more than the solvers with a sharp interface
(Fig.~\ref{fig:finalOverview}d). The error is
therefore caused neither by the smearing of the interface, as suggested by Wroniszewski et
al.\ \cite{Wroniszewski2014}, nor by its sharpness, as suggested by Qwist and
Christensen \cite{Qwist2023}. The present
paper confirms this. The error disappears when only the convection term is changed and the
interface advection is left as it is (Section~\ref{sec:final}), and the time lag that causes it is
also present in a single-phase flow (Section~\ref{sec:lag}). Figure~\ref{fig:problem} shows the error for \texttt{interIsoGoam}. Under the crest the velocity in the
top water cells exceeds the exact value, and the excess grows in time. Under the trough the velocity
is also too high in the top cells, and too low further down.
The error is therefore a spurious current in the water near the surface along the whole wave,
and below it there is a weaker return flow. Since the exact wave has no such current, it is a numerical error. In the present paper we show that it is caused by the momentum
convection term.

The consistency of mass and momentum transport is known to be essential at high density
ratios. Rudman \cite{Rudman1998} transported momentum with the mass fluxes of the volume
fraction advection, and this principle has been applied on staggered grids
\cite{Arrufat2020}, with level set and front tracking methods \cite{Liu2023} and with
isoAdvector \cite{Liu2024}. Liu et al.\ \cite{Liu2024} concluded that the mass flux computed
from the geometric fluxes of isoAdvector is consistent with Euler time stepping but not with
the Crank--Nicolson scheme, and that upwind momentum convection keeps the momentum transport
consistent with it. Symmetry-preserving
discretisations \cite{Verstappen2003,Veldman2019} make the convection operator neutral with
respect to kinetic energy, and Remmerswaal and Veldman \cite{Remmerswaal2022} combined this
with the geometric mass fluxes of a VoF method.

Explicit time integration of the convection term is common in incompressible flow solvers. In
the fractional-step method of Kim and Moin \cite{Kim1985}, widely used for direct and
large-eddy simulation of turbulence, the convection term is advanced with the explicit
Adams--Bashforth scheme and the viscous term with Crank--Nicolson. Vuorinen et al.\ \cite{Vuorinen2014}
implemented explicit Runge--Kutta projection methods for single-phase flow in \OF and found them
less dissipative than the standard PISO solver. The fractional-step wave
code NS3 \cite{Mayer1998,Christensen2006}, developed at DHI, also advances the convection term with
the Adams--Bashforth scheme, and the consistent VoF
methods of Rudman \cite{Rudman1998} and Popinet \cite{Popinet2009} also treat the convection
term explicitly. In these methods the convecting flux and the convected velocity belong to
the same time level. The \OF solvers instead treat convection implicitly. They convect the
velocity of the end of the time step with the mass flux of its start. This linearises the
convection term in the new velocity, so that it enters the momentum matrix and adds to its
diagonal. With an upwind-biased scheme the matrix then stays diagonally dominant, and the
time step is not limited by the convective stability limit of an explicit scheme
\cite{Ferziger2020}. The mass flux must be known to form the matrix, which is why it is taken
from the start of the step. For single-phase flow, Tukovi\'c et al.\ \cite{Tukovic2018}
noted that this old mass flux prevents second-order accuracy in
time, even with a second-order time scheme, and restored
second order by extrapolating the mass flux to the new time level. In a VoF solver the mass
flux is fixed by the interface advection and cannot be changed without breaking its
consistency with the volume fraction. The systematic mean momentum flux that the old mass flux
produces in an unsteady flow has, to our knowledge, not been reported.

The present paper adds two requirements to the consistency of mass and momentum transport.
The first concerns the \emph{time level} of the convected velocity, and the second the
\emph{convected face velocity at the interface}. We show how violating them causes the error
of the crest velocity, and how they can be met. The paper follows the sequence of changes to the
solver. Each section identifies an error and its cause, proposes a change and tests it.
Section~\ref{sec:method} summarises the discretisation. Section~\ref{sec:lag} shows that the
convected velocity lags the mass flux by a fraction of the time step, that this lag produces
the error of Fig.~\ref{fig:problem}, and that explicit convection removes it.
Section~\ref{sec:simple} shows with prescribed flows and with \texttt{interIsoGoam} that explicit
convection is consistent with the mass flux and with the gravity force of \texttt{interIsoGoam}.
Section~\ref{sec:interface} shows that explicit convection requires upwind convected face velocities at the
interface. Section~\ref{sec:wave} tests these changes on a progressive wave, where the
growth of the wave with explicit convection at larger time steps motivates a Lax--Wendroff correction
away from the interface. Section~\ref{sec:final} applies the resulting solver, \texttt{fluxinator}, to the wave study
of Part~I.
Section~\ref{sec:conclusion} gives the resulting recommendations.

The caption of each figure and table names a directory in the code repository of this paper
(\emph{Files}). It holds the script that makes the figure or table from the runs, and the
numbers it shows. Footnotes marked \emph{Data} give the file or run behind a number in the text.
Together they tie each result to the numerical experiment that produced it, so that the reader
can check that the claims of the paper rest on the data, and reproduce the data.
The results were computed with \OF v2512, and the code builds with \OF v2306 to v2606.

%% file: paper2/method.tex
\section{Momentum convection in a geometric VoF solver}
\label{sec:method}

We use the solver \texttt{interIsoGoam} of Part~I \selfcite{RoenbyPart1}. It solves the momentum and continuity
equations of an incompressible two-phase flow with density $\rho = \alpha\rho_1 +
(1-\alpha)\rho_2$ and modified pressure $p_{rgh} = p - \rho\,\grav\cdot\x$. Fluid 1 is the
heavier fluid, $\rho_1 > \rho_2$ (water and air in the two-phase test cases). The interface
is advected with isoAdvector \cite{Roenby2016,Scheufler2019}, and the gravity force uses the
geometric gravity model, which evaluates it with the density and the reconstructed interface at the
new time level.

\parhead{Mass flux.} In a time step from $t^n$ to $t^{n+1}$, isoAdvector computes the volume
$\Delta V_f$ of fluid 1 (water) that crosses each face $f$, given the volumetric face flux
$\phi_f^n$ at the start of the step. It then updates the volume fraction of cell $P$ (volume
$V_P$) as $\alpha_P^{n+1} = \alpha_P^n - \sum_f \Delta V_f/V_P$, where the sum runs over the
faces of $P$, oriented outward. The mass flux of the momentum equation, derived for isoAdvector by Roenby et al.\
\cite{Roenby2019}, is
\begin{equation}
  F_f = (\rho_1-\rho_2)\,\frac{\Delta V_f}{\Delta t} + \rho_2\,\phi_f^n .
  \label{eq:rhoPhi}
\end{equation}
Since $\sum_f\phi_f^n = 0$, the cell densities satisfy
\begin{equation}
  \frac{\rho_P^{n+1} - \rho_P^n}{\Delta t} + \frac{1}{V_P}\sum_f F_f = 0 .
  \label{eq:massConsistency}
\end{equation}
The isoAdvector update is the exact integral of the volume balance of the cell over the time
step, and the only approximation lies in the computation of $\Delta V_f$. Equation~\ref{eq:massConsistency} therefore has the two-level form of an Euler discretisation of the
continuity equation in which the mass flux (Eqn.~\ref{eq:rhoPhi}) is the average over the step. This
form implies no first-order truncation in time. This consistency of mass and volume transport was discussed by Rudman
\cite{Rudman1998} and, for isoAdvector, by Liu et al.\ \cite{Liu2024}.

\parhead{Momentum equation.} With the same two-level form (Euler time stepping) the momentum
equation of cell $P$ is
\begin{equation}
  \frac{\rho_P^{n+1}\U_P^{n+1} - \rho_P^n\U_P^n}{\Delta t}
  + \frac{1}{V_P}\sum_f F_f\,\U_f
  = \mathbf{R}_P ,
  \qquad
  \U_f = \theta\,\mathcal{I}_f(\U^{n+1}) + (1-\theta)\,\mathcal{I}_f(\U^n),
  \label{eq:momentum}
\end{equation}
where $\mathbf{R}_P$ contains the pressure gradient, the gravity force and the viscous term,
and $\mathcal{I}_f$ is the face interpolation of the convection scheme, here \texttt{limitedLinearV}
(the limited linear scheme of \OF for vectors) unless stated otherwise. We call $\U_f$, the velocity that the mass flux $F_f$ carries through face $f$, the
\emph{convected face velocity}. It is mostly a face interpolation of the cell velocities, but the
Lax--Wendroff correction of Section~\ref{sec:LW} adds a term from the velocity gradient. In all test cases of this paper
the fluids are inviscid and there is no surface tension, so that $\mathbf{R}_P$ contains only
the pressure gradient and the gravity or body force. The convected
velocity is taken at the new time level ($\theta = 1$, \emph{implicit} convection, hardcoded in \texttt{interFoam} and \texttt{interIsoFoam}), at the old time level ($\theta = 0$, \emph{explicit} convection), or in
between (e.g.\ $\theta = \tfrac12$, Crank--Nicolson). The mass flux $F_f$ is the same in all 
cases. Equation~\ref{eq:momentum} describes a step with one outer corrector, which is the setting
used throughout this paper. The mass flux is then computed once from the face flux $\phi^n$ of
the start of the step, and the pressure equation makes the new face flux $\phi^{n+1}$
divergence free. With more than one outer corrector, \texttt{interIsoGoam} and \texttt{fluxinator}
by default centre the step in time instead: the interface advection, the convected velocity and the
gravity force each take the mean of their values at the start and at the latest end of the step.
This mode is not tested in this paper. In all runs of
this paper the \texttt{ddtCorr} term of \OF, which adds a part of the old face flux to the
predicted face flux, is switched off. Part~I \selfcite{RoenbyPart1} shows that it adds a damping
that decreases only slowly, if at all, as the time step is reduced.
Except for the \texttt{fluxinator} runs of Section~\ref{sec:final}, the runs use the solver \texttt{interIsoGoam} of Part~I (\datasource{applications/solvers/interIsoGoam}).
In it the weight $\theta$ of Eqn.~\ref{eq:momentum} is set by \texttt{convectionNewTimeWeight}, an
entry we introduce in the \texttt{PIMPLE} dictionary of \texttt{system/fvSolution}. Its default is 1
(implicit convection, the solver of Part~I). The convection scheme, and thereby $\mathcal{I}_f$, is
set by the entry \texttt{div(rhoPhi,U)} in the \texttt{divSchemes} dictionary of
\texttt{system/fvSchemes}. The prescribed flows of Section~\ref{sec:simple} use a separate test application.

%% file: paper2/timeLag.tex
\section{The time lag between mass flux and convected velocity}
\label{sec:lag}

In Eqn.~\ref{eq:momentum} the mass flux, $F_f$, belongs to $t^n$, while the convected velocity belongs to
$t^n + \theta\Delta t$. Consider the flux of horizontal momentum through a horizontal face. It
is the mass flux, which contains the vertical velocity $w$, times the convected horizontal
velocity $u$. Averaged over the horizontal direction, the discrete flux therefore
approximates
\begin{equation}
  \rho\,\langle w(t)\,u(t+\theta\Delta t)\rangle
  = \rho\,\langle w u\rangle
  + \theta\Delta t\,\rho\,\Bigl\langle w\,\frac{\partial u}{\partial t}\Bigr\rangle + \mathcal{O}(\Delta t^2) .
  \label{eq:lagFlux}
\end{equation}
The second term is a spurious Reynolds stress. It vanishes in a steady flow, but not in an unsteady
flow in which $w$ and $\partial u/\partial t$ are correlated, such as a travelling or oscillating
flow pattern. It is
proportional to the time step, vanishes for explicit convection ($\theta = 0$) and is halved by
Crank--Nicolson. The lag moves momentum, but it does not create it.

\begin{figure}[tbp]
\centering
\includegraphics[width=\linewidth]{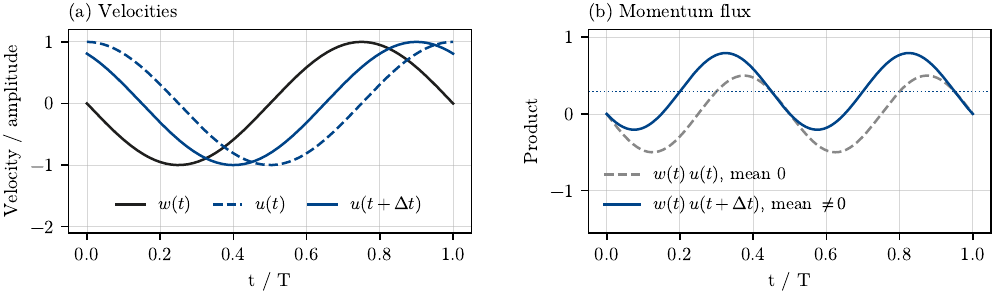}
\caption{The time lag at a point below the trough of a progressive wave. (a) $u$ and $w$
are 90$^\circ$ out of phase. (b) Their product has zero mean, but pairing the mass flux at $t$
with the velocity at $t + \Delta t$ gives a positive mean (here $\Delta t = T/10$ for the wave period $T$).
\figdir{fig_mechanismLag}.}
\label{fig:mechanismLag}
\end{figure}

\parhead{The lag in a progressive wave.} Below the trough of a progressive wave, $u$ and $w$
are 90$^\circ$ out of phase, and the exact flux of horizontal momentum through a horizontal plane,
averaged over a wavelength, is zero. This does not require linear theory, since in a wave
that is symmetric about its crest every harmonic of $u$ is a cosine and every harmonic of $w$
a sine of $kx-\omega t$, where $k$ is the wave number and $\omega$ the angular frequency. The lag term does not vanish: in a wave of permanent form with
celerity $c$, $\partial u/\partial t = -c\,\partial u/\partial x$, and with continuity,
$\partial u/\partial x = -\partial w/\partial z$, the lag term of Eqn.~\ref{eq:lagFlux} becomes
\begin{equation}
  \theta\Delta t\,\rho\,\Bigl\langle w\,\frac{\partial u}{\partial t}\Bigr\rangle
  = \tfrac12\,\theta\Delta t\,\rho\,c\,\frac{\mathrm{d}\langle w^2\rangle}{\mathrm{d}z} .
  \label{eq:lagFluxWave}
\end{equation}
This is an upward flux of horizontal momentum where $\langle w^2\rangle$ increases with height,
as it does everywhere below the trough (Fig.~\ref{fig:mechanismLag}).
For $u = \hat u\cos(kx-\omega t)$ and $w = \hat w\sin(kx-\omega t)$ of linear theory it
is $\tfrac12\theta\Delta t\,\rho\,\omega\,\hat u\,\hat w$. We consider heights below
the trough because the flux is an average over a horizontal plane, that is, over all phases of
the wave. It is not a profile at one phase, like the velocity under the crest that we use
elsewhere in this paper. Under the crest $w = 0$, so the column there carries no vertical flux.
The flux arises between crest and trough, where $w$ is large. Below the trough the plane lies
in water along its whole length, and $u$ and $w$ vary smoothly in time at every point of it.
Above the trough each point is alternately in water and in air, $u$ jumps when the interface
passes, and the expansion (Eqn.~\ref{eq:lagFlux}) does not hold there. The upward flux cannot leave the water through the surface, so the momentum
accumulates in the top water cells as a spurious surface current that grows in time, and a
compensating return flow develops below. The horizontal velocity under the crest then exceeds the
exact velocity near the surface and falls short of it further down. This is the error of
Fig.~\ref{fig:problem}.

\begin{figure}[tbp]
\centering
\includegraphics[width=\linewidth]{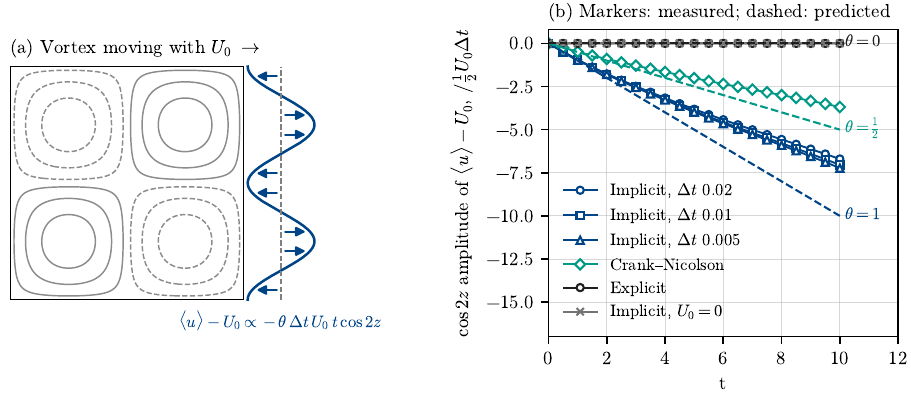}
\caption{The time lag in a single-phase flow: Taylor--Green vortex moving with $U_0$.
(a) Streamlines, dashed where the flow is clockwise, and the predicted spurious mean flow. (b) Amplitude of
the spurious mean flow $\langle u\rangle - U_0 \propto \cos 2z$, divided by $\tfrac12 U_0\Delta t$,
against time (markers), and the prediction $-\theta t$ (dashed).
\figdir{fig_lagTGV}.}
\label{fig:lagTGV}
\end{figure}

\parhead{A single-phase test.} The mechanism does not depend on the interface. The
Taylor--Green vortex $u = U_0 + \sin(x - U_0 t)\cos z$, $w = -\cos(x - U_0 t)\sin z$ in a
doubly periodic box is an exact solution of the Euler equations whose horizontal mean
velocity stays $U_0$. Its pattern moves with $U_0$, so $\partial u/\partial t = -U_0\,\partial
u/\partial x$ is correlated with $w$, and the lag predicts the spurious mean flow
$\langle u\rangle - U_0 = -\tfrac12\theta\Delta t\,U_0\,t\cos 2z$ (Fig.~\ref{fig:lagTGV}a).
We ran \texttt{interIsoGoam} on a $64\times64$ mesh with $\alpha = 1$ everywhere. The measured amplitude
is proportional to the time step, halved by Crank--Nicolson, and absent with explicit
convection and without translation (Fig.~\ref{fig:lagTGV}b). It follows the prediction at
first and falls to about 70\,\% of it at $t = 10$, as the vortex loses energy to the numerical
diffusion of implicit convection. Total momentum is conserved to round-off. With $U_0 = 0$
the effect vanishes, which may be why the usual stationary Taylor--Green test has not
revealed it.

Explicit convection removes the lag. Before testing it in the wave
(Section~\ref{sec:wave}), we check in Sections~\ref{sec:simple} and~\ref{sec:interface} that
explicit convection is consistent with the mass flux and the force, and what it requires of
the convected face velocity at the interface.

%% file: paper2/simpleFlows.tex
\section{Consistency of explicit convection with the mass flux and the force}
\label{sec:simple}

Changing the time level of the convected velocity must not break the consistency of momentum
and mass transport. Two flows should be reproduced exactly by the discrete equations: a
uniform velocity and a uniform acceleration. We derive the conditions for both and test them
with prescribed flows, in which only the momentum equation is solved, and with \texttt{interIsoGoam}.

\subsection{Uniform velocity: the time scheme}
\label{sec:uniform}

Let the velocity be uniform, $\U = \U_0$, with no force. Every face interpolation that
reproduces a constant gives $\U_f = \U_0$, and the left-hand side of Eqn.~\ref{eq:momentum}
becomes $\U_0$ times the left-hand side of Eqn.~\ref{eq:massConsistency}, which vanishes. The
uniform velocity is therefore an exact discrete solution for every convection scheme and
every $\theta$. With the second-order backward scheme the time derivative becomes
$(3\rho^{n+1}\U^{n+1} - 4\rho^n\U^n + \rho^{n-1}\U^{n-1})/(2\Delta t)$, and the residual is
\begin{equation}
  \U_0\,\frac{\rho^{n+1} - 2\rho^n + \rho^{n-1}}{2\Delta t} ,
  \label{eq:backwardResidual}
\end{equation}
which is non-zero in every cell whose volume fraction does not change linearly in time, that
is in the cells the interface passes through. Divided by a small cell density, it gives a
large velocity error in those cells. The mass flux (Eqn.~\ref{eq:rhoPhi}) of the step is therefore
consistent with Euler time stepping but not with the backward scheme. At a constant time step,
the backward density difference equals the divergence of $\tfrac32 F_f - \tfrac12 F_f^{-}$,
where $F_f^{-}$ is the mass flux of the previous step. A backward scheme that convects momentum
with this combination would preserve a uniform velocity. We do not pursue it here.

\subsection{Uniform acceleration: the time level of the force}
\label{sec:timeLevel}

Let the velocity be uniform but time dependent, with acceleration
$\mathbf{a} = \mathrm{d}\U/\mathrm{d}t$. The viscous term vanishes, and $\mathbf{R}_P$ in
Eqn.~\ref{eq:momentum} is the force per unit volume that drives the acceleration. Inserting
$\U_f = \theta\U^{n+1} + (1-\theta)\U^n$ in Eqn.~\ref{eq:momentum} and using
Eqn.~\ref{eq:massConsistency} gives
\begin{equation}
  \bigl[\theta\rho_P^n + (1-\theta)\rho_P^{n+1}\bigr]\,\frac{\U_P^{n+1} - \U_P^n}{\Delta t}
  = \mathbf{R}_P .
  \label{eq:timeLevel}
\end{equation}
The exact solution, $\U^{n+1} - \U^n = \mathbf{a}\,\Delta t$, is therefore reproduced only if
$\mathbf{R}_P = \bigl[\theta\rho^n + (1-\theta)\rho^{n+1}\bigr]\mathbf{a}$, that is, if the
force is built from the old density for implicit convection and from the new density for explicit
convection. In cells that the interface
passes through, $\rho^n$ and $\rho^{n+1}$ can differ by a factor of up to the density ratio
$\rho_1/\rho_2$. The gravity
force of \texttt{interFoam} and \texttt{interIsoFoam}, $-gh_f\,\nabla_f\rho^{n+1}$, that of the geometric model of \texttt{interIsoGoam}, and body forces of the form
$\rho^{n+1}\mathbf{a}$ use the new density. They are consistent with explicit convection,
and inconsistent with the implicit convection hardcoded in \texttt{interFoam} and \texttt{interIsoFoam}.

\subsection{Tests}
\label{sec:consistencyTests}

\parhead{Prescribed flows.} To test the convection term in
isolation we prescribe the velocity and solve only the momentum equation, without pressure. A
disc of water (radius 0.15\,m, density 1000\,kg/m$^3$) in air (1\,kg/m$^3$) is placed in a
two-dimensional box of $1\times1$\,m with $50\times50$ cells. The flow is defined by a stream
function $\psi$, and the face fluxes are obtained by integrating $\psi$ along the edges of
each face, so that they are divergence free to round-off on any mesh. Each step advances
$\alpha$ with isoAdvector (interface reconstruction \texttt{plicRDF}) and then solves
\begin{equation}
  \frac{\rho^{n+1}\U^{n+1} - \rho^n\U^n}{\Delta t} + \frac{1}{V}\sum_f F_f\,\U_f
  = \rho^*\,\mathbf{a} ,
  \label{eq:rig}
\end{equation}
whose right-hand side is $\mathbf{R}_P$ of Eqn.~\ref{eq:momentum}, written as
$\rho^*\mathbf{a}$. Here $\mathbf{a} = \partial_t\mathbf{u} + (\mathbf{u}\cdot\nabla)\mathbf{u}$ is the
exact acceleration at the cell centre, which a pressure gradient or body force would have to
supply, and $\rho^*$ is $\rho^{n+1}$ unless stated otherwise. The error is the difference
from the prescribed velocity at the cell centre. In \emph{one-step} mode each step starts
from the exact velocity, so the error is that of a single step. In \emph{carried} mode the
computed velocity is the old velocity of the next step, so the errors accumulate. In both
modes the interface advection uses only the prescribed flow, since isoAdvector receives the
prescribed face fluxes and the prescribed cell velocities, which it uses to move the reconstructed
interface within the step. The mass flux therefore comes from the prescribed flow, and the computed
velocity is convected, but it does not move the fluid. We use two flows: in the first the disc
moves with the uniform velocity $\U_0 = (0.2, 0.05)$\,m/s,
and in the second it is uniformly accelerated from $\U_0 = (0.1, 0.05)$\,m/s with
$\mathbf{a} = (0.5, 0)$\,m/s$^2$. The time step is 0.005\,s and the duration 1\,s.
Table~\ref{tab:rigUniform} confirms both analyses. A uniform velocity
is preserved to round-off with Euler time stepping, while the backward scheme produces errors
of 2.4\,m/s in a flow of 0.2\,m/s. Under uniform acceleration only two of the five combinations are exact, namely implicit convection with the force built from $\rho^n$ and explicit convection with
the force built from $\rho^{n+1}$.

\begin{table}[tbp]
\centering
\caption{Prescribed uniform flows: largest velocity error [m/s] over all time steps.
Uniform: $\U_0 = (0.2, 0.05)$\,m/s. Uniform acceleration: $\U_0 = (0.1, 0.05)$\,m/s,
$\mathbf{a} = (0.5, 0)$\,m/s$^2$. Face interpolation \texttt{limitedLinearV}. All face interpolations
give the same values for these flows.
\figdir{tab_rigUniform}.}
\label{tab:rigUniform}
\small
\begin{tabular}{@{}lccc@{}}
\toprule
 & uniform & \multicolumn{2}{c}{uniform acceleration} \\
\cmidrule(l){3-4}
Convection, time scheme, force density & one step & one step & carried \\
\midrule
implicit, backward, $\rho^{n+1}$ & $2.4$ & $15$ & $23$ \\
implicit, Euler, $\rho^n$     & $1.7\times10^{-16}$ & $3.9\times10^{-13}$ & $5.9\times10^{-14}$ \\
implicit, Euler, $\rho^{n+1}$ & $1.7\times10^{-16}$ & $5.0\times10^{-3}$ & $5.0\times10^{-2}$ \\
explicit, Euler, $\rho^n$     & $1.7\times10^{-15}$ & $2.1\times10^{-1}$ & $3.0$ \\
explicit, Euler, $\rho^{n+1}$ & $1.7\times10^{-15}$ & $1.4\times10^{-14}$ & $7.0\times10^{-14}$ \\
\bottomrule
\end{tabular}
\end{table}

\begin{figure}[tbp]
\centering
\includegraphics[width=0.8\linewidth]{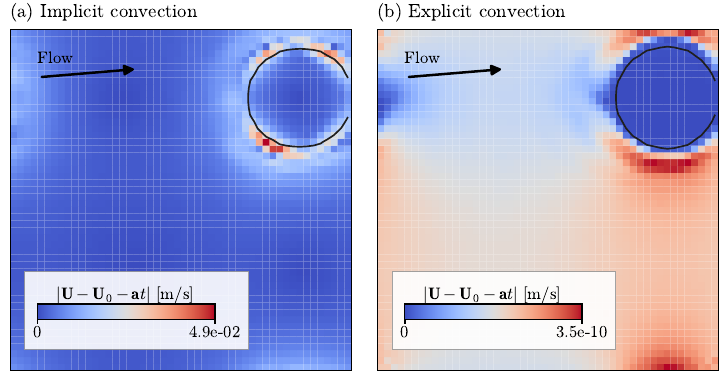}
\caption{\texttt{interIsoGoam}, disc in a periodic box under uniform acceleration: deviation from the
exact velocity at $t = 1$\,s with (a) implicit and (b) explicit convection (\texttt{limitedLinearV}, force
built with $\rho^{n+1}$). The black line is the contour $\alpha = 0.5$, and the arrow shows the
direction of the flow. Each panel has its own colour scale.
\figdir{fig_fullSolver}.}
\label{fig:fullSolver}
\end{figure}

\parhead{Pressure--velocity coupling.} The prescribed-flow tests have no pressure. To include the pressure--velocity coupling we
run \texttt{interIsoGoam} on the \OF tutorial \texttt{discInConstantFlowCyclicBCs}, which is a water disc of radius 0.15\,m in air in a fully periodic box of $1\times1$\,m with $50\times50$ cells. The interface
is reconstructed with \texttt{plicRDF}. There is no gravity. The initial velocity is
$\U_0 = (0.1, 0.05)$\,m/s, and a uniform body force $\rho^*\mathbf{a}$ with
$\mathbf{a} = (0.5, 0)$\,m/s$^2$ drives the flow. The exact solution is $\U = \U_0 + \mathbf{a}t$ with a
uniform pressure. With implicit convection and the force built from $\rho^{n+1}$, which is
the combination hardcoded in these solvers, spurious velocities of 0.05\,m/s and pressure
variations of 1.7\,Pa appear around the disc (Tab.~\ref{tab:fullSolver},
Fig.~\ref{fig:fullSolver}). With explicit convection and the same force, the solution is
exact to the tolerance of the pressure solver. Without the force, a uniform velocity is
preserved to round-off with both implicit and explicit convection.\footnote{\datanote{papers/part2/tab_fullSolver/fullSolver.txt},
cases \texttt{noForce\_*}.} Since the gravity force of \texttt{interIsoGoam} is built with the new density,
explicit convection is the consistent choice. Building the force with the old density instead would
make implicit convection consistent with it, but would leave the time lag of Section~\ref{sec:lag}.

\begin{table}[tbp]
\centering
\caption{\texttt{interIsoGoam}, disc in a periodic box under uniform acceleration: largest deviation
[m/s] from the exact velocity $\U_0 + \mathbf{a}t$ over 1\,s, in water ($\alpha > 0.99$),
in cells with $0.01 \le \alpha \le 0.99$ and in air ($\alpha < 0.01$), and the range of $p_{rgh}$, which is uniform in the exact solution. Face
interpolation \texttt{limitedLinearV}.
\figdir{tab_fullSolver}.}
\label{tab:fullSolver}
\small
\begin{tabular}{@{}lcccc@{}}
\toprule
Convection, force density & water & interface & air & $p_{rgh}$ range [Pa] \\
\midrule
implicit, $\rho^{n}$   & $9.4\times10^{-5}$ & $1.0\times10^{-4}$ & $3.3\times10^{-4}$ & $1.2\times10^{-2}$ \\
implicit, $\rho^{n+1}$ & $2.4\times10^{-2}$ & $4.0\times10^{-2}$ & $4.9\times10^{-2}$ & $1.7$ \\
explicit, $\rho^{n}$   & $2.2\times10^{-2}$ & $1.2\times10^{-1}$ & $1.4\times10^{-1}$ & $3.4$ \\
explicit, $\rho^{n+1}$ & $6.2\times10^{-12}$ & $4.6\times10^{-11}$ & $3.5\times10^{-10}$ & $1.8\times10^{-9}$ \\
\bottomrule
\end{tabular}
\end{table}

%% file: paper2/interface.tex
\section{Explicit convection at the interface}
\label{sec:interface}

The uniform flows of Section~\ref{sec:simple} do not depend on the face interpolation of the cell-centred velocity field. When
the velocity varies across the interface, the interpolation does matter, and more so for
explicit than for implicit convection, as we now show.

\subsection{\texorpdfstring{The convected face velocity at the interface}{The convected face velocity at the interface}}
\label{sec:faceValues}

Without pressure and body forces, one explicit step (Eqn.~\ref{eq:momentum}) changes the momentum of a cell $P$ by the
momentum that crosses its faces. With the cell mass $m_P = \rho_P V_P$ and $F_f$ counted positive
out of $P$,
\begin{equation}
  m_P^{n+1}\U_P^{n+1} = m_P^n\U_P^n - \Delta t\sum_f F_f\,\U_f ,
  \qquad
  m_P^{n+1} = m_P^n - \Delta t\,F_{\mathrm{out}} + \Delta t\,F_{\mathrm{in}} ,
  \label{eq:explicitStep}
\end{equation}
where the mass balance is Eqn.~\ref{eq:massConsistency}, and $\Delta t\,F_{\mathrm{out}}$ and
$\Delta t\,F_{\mathrm{in}}$ are the masses that leave and enter $P$ in the step. With upwind convected face velocities, the momentum leaves with $\U_P^n$ and enters with the velocity of the cell it comes from:
\begin{equation}
  \U_P^{n+1} = \frac{m_P^n - \Delta t\,F_{\mathrm{out}}}{m_P^{n+1}}\,\U_P^n
  + \sum_{f\,\in\,\mathrm{in}} \frac{\Delta t\,|F_f|}{m_P^{n+1}}\,\U_{N(f)}^n ,
  \label{eq:upwindAverage}
\end{equation}
where the sum runs over the faces through which mass enters $P$.
Since $F_f$ is counted positive out of $P$, these are the faces with $F_f < 0$, and
$\Delta t\,|F_f|$ is the mass that enters through face $f$. $N(f)$ is the cell on the
other side of face $f$. We call the coefficients of the old velocities in
Eqn.~\ref{eq:upwindAverage} the weights of the update. By the mass balance in
Eqn.~\ref{eq:explicitStep} they sum to one, and they are non-negative as long as the mass that leaves in a step does
not exceed the mass in the cell at its start, $\Delta t\,F_{\mathrm{out}} \le m_P^n$. The new
velocity is then a weighted average of old velocities, whatever the density ratio, and it cannot leave the
range of the old velocities of $P$ and its upwind neighbours.

\begin{figure}[tbp]
\centering
\includegraphics{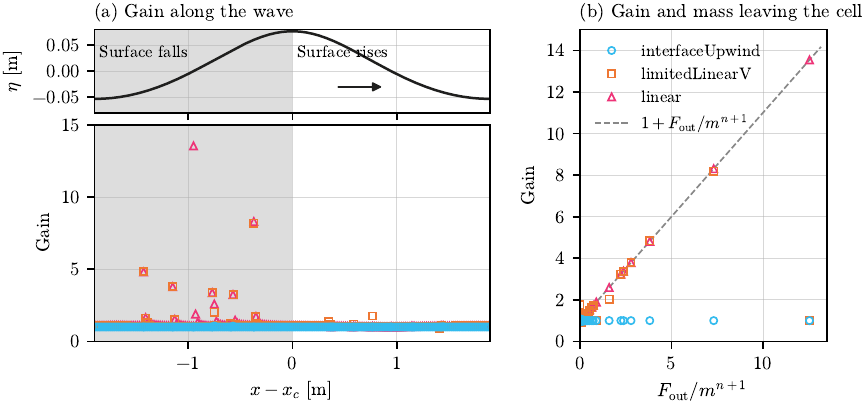}
\caption{Gain (Eqn.~\ref{eq:gain}) of one explicit convection step in the interface cells of a linear
two-phase wave ($190\times40$ cells, $\Delta t = 0.005$\,s, exact mass fluxes and exact mass-weighted
velocities at $t^n$). (a) Surface elevation and gain along the wave. The crest is at $x = x_c$ and
travels to the right, and the surface falls in the shaded half. (b) Gain against the mass that leaves
the cell in the step, relative to the mass after the step. The dashed line is Eqn.~\ref{eq:gain}.
\figdir{fig_mechanismFace}.}
\label{fig:mechanismFace}
\end{figure}

A convected face velocity that takes part of its value from the downwind cell loses this property. With the
linear interpolation $\U_f = \tfrac12(\U_P + \U_N)$, each downwind neighbour gets the weight
$-\Delta t\,F_f/(2m_P^{n+1})$. As long as the weight of $P$ itself stays non-negative, the sum of
the absolute weights, which we call the gain of the cell, is
\begin{equation}
  G_P = \sum_k \left|\frac{\partial \U_P^{n+1}}{\partial \U_k^n}\right|
      = 1 + \frac{\Delta t\,F_{\mathrm{out}}}{m_P^{n+1}} ,
  \label{eq:gain}
\end{equation}
where $k$ runs over $P$ and its face neighbours, which are the cells whose old
velocities affect $\U_P^{n+1}$. $G_P$ is the row sum of the absolute coefficients of the update,
and its largest value over all cells is the maximum norm of the matrix that maps the old velocities to
the new ones. $G_P = 1$ with non-negative weights is the positive-coefficient condition of
monotone schemes \cite{Spekreijse1987,Jameson1995}. The gain matters because the old velocities are never exact. Interface cells carry errors from
earlier steps and from the pressure correction. For upwind and linear interpolation the weights depend
only on the mass fluxes, and an error $\delta\U_k^n$ in the old velocities changes the new one by at most
\begin{equation}
  |\delta\U_P^{n+1}| \le \sum_k |w_k|\,|\delta\U_k^n| \le G_P \max_k |\delta\U_k^n| ,
  \label{eq:gainBound}
\end{equation}
where $w_k$ are the weights. With $G_P = 1$ convection cannot make an error grow, but with $G_P > 1$
it can. As an example, take a cell that loses its water through a single face and receives nothing
in the step. Suppose that the mass that leaves is 10 times the mass that remains, which is somewhat less than in the cell with
the largest gain in Fig.~\ref{fig:mechanismFace}. With upwind interpolation $\U_P^{n+1} = \U_P^n$, so the
remaining air keeps the velocity of the cell. With linear interpolation,
\begin{equation}
  \U_P^{n+1} = 6\,\U_P^n - 5\,\U_N^n .
  \label{eq:gainExample}
\end{equation}
The weights still sum to one, so a uniform velocity is kept, but a difference of 0.1\,m/s between
$\U_P^n$ and $\U_N^n$ puts $\U_P^{n+1}$ 0.5\,m/s outside both. The gain is large where a cell
loses most of its mass in a step. The water leaves with a face velocity partly taken from the
downwind cell, and the momentum it takes out is divided by the little mass that remains, mostly air. Since the mass that remains can be as small as that of the air, the gain
can reach about the Courant number times the density ratio. A small time step therefore does not keep it
close to one.

Figure~\ref{fig:mechanismFace} shows where this happens in a wave. We evaluated one explicit step in
a linear two-phase wave with wavelength 3.79\,m,
height 0.125\,m and water depth 0.4\,m, with 0.4\,m of air above, on $190\times40$ cells with
$\Delta t = 0.005$\,s. The mass fluxes and the mass-weighted cell velocities are exact, and we
computed the gain of every interface cell.\footnote{Model:
\datasource{papers/part2/crestFaces.py}. The surface is a streamline of the
water flow in the frame of the wave, so the mass fluxes follow exactly from the stream functions.} Large gains occur only
where the surface falls, that is, behind the crest in cells that the water leaves. There, linear convected face velocities
give gains up to 13.5, exactly Eqn.~\ref{eq:gain}, and \texttt{limitedLinearV} up to 8.2. The
condition $\Delta t\,F_{\mathrm{out}} \le m_P^n$ holds in all cells (largest ratio 0.93).

The upwind convected face velocity is not exact either. With one velocity per cell, the velocity of an interface
cell is a mass-weighted mix of water and air velocity, and upwind interpolation gives the water that leaves
the cell this mixed velocity. In the same wave step, this causes errors of up to 0.24\,m/s in single cells that are being
emptied, whereas \texttt{limitedLinearV} gives 0.13\,m/s. The advantage of upwind interpolation at the
interface is therefore not a smaller error in one step, but that the step stays a weighted
average, so that errors from earlier steps cannot grow.\datafn{papers/part2/data/crestSchemes_dt0.005.txt}

We propose the new scheme \emph{\texttt{interfaceUpwind}}. It takes the upwind value on every face of an
interface cell ($\epsilon < \alpha < 1-\epsilon$, with $\epsilon$ the tolerance \texttt{surfCellTol} that defines the interface cells of isoAdvector) and on every face between
a water cell and an air cell. On faces between two cells of the same pure phase it uses an
underlying scheme $\mathcal{I}_f$, here \texttt{limitedLinearV}. For a face $f$ between the cells
$P$ and $N$,
\begin{equation}
  \U_f =
  \begin{cases}
    \mathcal{I}_f(\U) & \text{if } \alpha_P, \alpha_N \le \epsilon
                        \text{ or } \alpha_P, \alpha_N \ge 1-\epsilon,\\[2pt]
    \U_P & \text{otherwise, if } F_f \ge 0,\\[2pt]
    \U_N & \text{otherwise, if } F_f < 0,
  \end{cases}
  \label{eq:interfaceUpwind}
\end{equation}
with $F_f$ counted positive from $P$ to $N$.
Its implementation is in \datasource{src/interfaceSchemes}. In the wave step of
Fig.~\ref{fig:mechanismFace} its gain is one, except in a few cells that are pure water or pure
air at $t^n$ and become mixed in the step, where \texttt{limitedLinearV} is still used on faces to
pure neighbours (at most 1.07). With implicit
convection the outflow term of Eqn.~\ref{eq:explicitStep} is taken at $t^{n+1}$ and enters the diagonal
of the momentum matrix instead of being subtracted from the old momentum, so the amplification by the gain (Eqn.~\ref{eq:gain}) does not occur.

\subsection{Tests with non-uniform flows}
\label{sec:nonUniform}

\begin{figure}[tbp]
\centering
\includegraphics[width=0.67\linewidth]{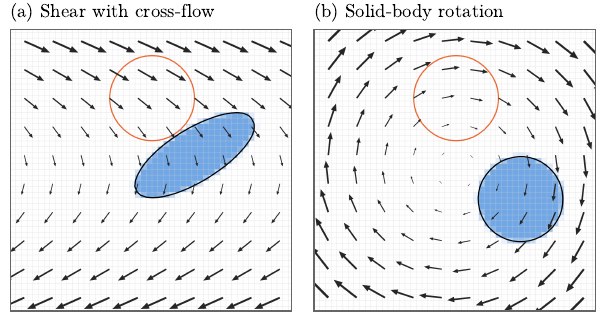}
\caption{Prescribed non-uniform flows at $t = 1$\,s: volume fraction on the
$50\times50$ mesh (white air, blue water), the interface $\alpha = 0.5$ at $t = 0$ (orange)
and $t = 1$\,s (black), and the prescribed velocity (arrows at every fifth cell, with lengths of
0.2 and 0.09\,m per m/s). ParaView snapshots.
\figdir{fig_rigCasesFlows}.}
\label{fig:rigCasesFlows}
\end{figure}

\begin{figure}[tbp]
\centering
\includegraphics[width=\linewidth]{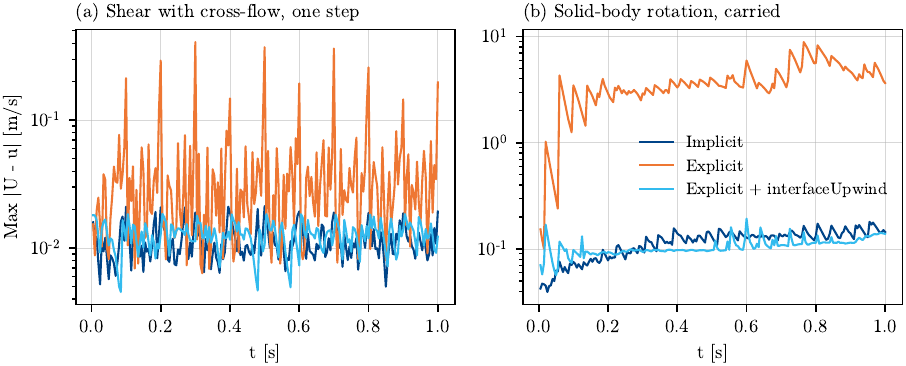}
\caption{Prescribed flows: largest velocity error in each time step. (a) Shear with cross-flow,
one-step mode. (b) Solid-body rotation, carried mode.
\figdir{fig_rigErrors}.}
\label{fig:rigErrors}
\end{figure}

We return to the prescribed flows of
Section~\ref{sec:consistencyTests}, now with two flows whose velocity varies across the
interface (Fig.~\ref{fig:rigCasesFlows}): a shear flow with a cross-flow,
$\mathbf{u} = ((z - 0.5)\,\mathrm{s}^{-1},\,-0.2\,\mathrm{m/s})$, and solid-body rotation with
$\Omega = 2$\,rad/s about the box centre. In both the disc is centred at $(0.5, 0.75)$\,m at
$t = 0$. The time step is again 0.005\,s and the duration 1\,s. Explicit convection with
\texttt{limitedLinearV} or linear convected face velocities gives errors of the order of the flow speed or larger, and they occur in single steps
in a few interface cells (Tab.~\ref{tab:rigFaces}, Fig.~\ref{fig:rigErrors}). Implicit convection
with \texttt{limitedLinearV}, for which the amplification by the gain does not occur, gives errors 20 to 50
times smaller. With
\texttt{interfaceUpwind} the errors of explicit convection are of the same order as those of implicit
convection. Under mesh refinement, the one-step errors of implicit convection and of explicit
convection with \texttt{interfaceUpwind} decrease roughly in proportion to the cell size. Those of explicit
convection with \texttt{limitedLinearV} do not decrease consistently, and in the carried rotation
they remain of order 10\,m/s (Tab.~\ref{tab:rigConvergence}). Part of the remaining error
in interface cells is a property of the reference. For a linearly varying velocity, the
mass-averaged velocity of a partly filled cell differs from the velocity at its centre by up
to $|\nabla\mathbf{u}|\,h/2$, which is 0.01\,m/s for the shear flow on the $50\times50$ mesh.
In the uniform acceleration of Section~\ref{sec:consistencyTests}, explicit convection with
\texttt{interfaceUpwind} remains exact in \texttt{interIsoGoam}.\footnote{\datanote{papers/part2/tab_fullSolver/fullSolver.txt},
case \texttt{explicitIU\_rhoNew}.}

\begin{table}[tbp]
\centering
\caption{Prescribed non-uniform flows: largest velocity error [m/s] over all time steps.
Flow speeds at the disc are up to about 0.45\,m/s (shear) and 0.8\,m/s (rotation).
\figdir{tab_rigFaces}.}
\label{tab:rigFaces}
\small
\begin{tabular}{@{}lccccc@{}}
\toprule
 & \multicolumn{2}{c}{shear with cross-flow} & \multicolumn{2}{c}{solid-body rotation} & \\
\cmidrule(lr){2-3}\cmidrule(lr){4-5}\cmidrule(l){6-6}
Convection, face interpolation & one step & carried & one step & carried & \\
\midrule
implicit, limitedLinearV        & $0.021$ & $0.043$ & $0.11$ & $0.18$ & \\
explicit, limitedLinearV        & $0.41$  & $1.1$  & $2.2$  & $8.9$  & \\
explicit, linear                & $0.41$  & $2.3$   & $2.2$  & $390$  & \\
explicit, interfaceUpwind       & $0.019$ & $0.055$ & $0.36$ & $0.19$ & \\
\bottomrule
\end{tabular}
\end{table}

\begin{table}[tbp]
\centering
\caption{Prescribed flows, mesh refinement with the time step proportional to the cell size:
largest velocity error [m/s] over all time steps.
\figdir{tab_rigConvergence}.}
\label{tab:rigConvergence}
\small
\begin{tabular}{@{}lcccccc@{}}
\toprule
 & \multicolumn{3}{c}{shear, one step} & \multicolumn{3}{c}{rotation, carried} \\
\cmidrule(lr){2-4}\cmidrule(l){5-7}
Convection, face interpolation & $25^2$ & $50^2$ & $100^2$ & $25^2$ & $50^2$ & $100^2$ \\
\midrule
implicit, limitedLinearV   & $0.041$ & $0.021$ & $0.011$  & $0.27$ & $0.18$ & $0.13$ \\
explicit, limitedLinearV   & $0.32$  & $0.41$  & $0.21$   & $16$   & $8.9$  & $7.2$ \\
explicit, interfaceUpwind  & $0.036$ & $0.019$ & $0.011$ & $0.81$ & $0.19$ & $0.17$ \\
\bottomrule
\end{tabular}
\end{table}

\subsection{Relation to the analysis of Liu et al.}
\label{sec:liu}

Liu et al.\ \cite{Liu2024} analysed the same mass flux (Eqn.~\ref{eq:rhoPhi}) and drew conclusions
for the convection term, some of which the present results confirm and some of which they revise.
\begin{itemize}
\item \emph{Time scheme.} They showed that the mass flux, which gives the mass
that crosses a face during the whole time step, is
consistent with Euler time stepping but not with Crank--Nicolson time stepping, which combines
the mass fluxes of two time levels (their Section~3.1). This agrees with Section~\ref{sec:uniform}, which shows that the time
difference must match the mass flux of the step. Their objection to Crank--Nicolson does not apply to the Crank--Nicolson
convection of Eqn.~\ref{eq:momentum}, which weights the convected velocity but keeps a single mass
flux. They also listed the second-order backward scheme as consistent, because it uses a single
mass flux, but did not test it. Its density difference does not match the mass flux of the
step, however (Eqn.~\ref{eq:backwardResidual}), and it does not preserve a uniform velocity
(Tab.~\ref{tab:rigUniform}).
\item \emph{Mechanism of the errors of the convected face velocity.} They attributed the errors of limited and
blended convection schemes to a modification of the mass flux (their Sections~3.1 and~5). The convection scheme,
however, interpolates the convected velocity, and the mass flux is the same for every scheme. In
their runs that reach the final time, the mass error is of order $10^{-10}$ or smaller with
every scheme, and it grows only in the runs that end in failure (their Fig.~4). The errors therefore come from the
convected face velocity, not from the mass flux. With explicit convection, convected face velocities
that are not upwind amplify velocity differences in cells that lose most of their mass in a step
(Section~\ref{sec:faceValues}). Their runs use implicit convection, for which this amplification
does not occur, and we have not identified the mechanism there.
\item \emph{Where upwinding is needed.} They concluded that upwind momentum convection is consistent
with the geometric advection of the volume fraction, and used it in the whole domain (their
Section~5). Upwind
interpolation is bounded, but only first-order accurate and numerically diffusive
\cite[Sect.~4.4.1]{Ferziger2020}, so when it is used everywhere it adds this diffusion to the whole
flow. Upwind convected face velocities are needed only at the interface. With \texttt{interfaceUpwind}, which keeps
\texttt{limitedLinearV} away from the interface, the errors are of the same order as those of implicit convection
(Tabs.~\ref{tab:rigFaces} and~\ref{tab:rigConvergence}).
\item \emph{Time level of the force.} They concluded that Euler time stepping with upwind
convection guarantees consistency (their Section~3.4). With implicit convection this holds without forces, but
not when a force is built with the new density, as the gravity force of \texttt{interIsoGoam} is. Under
a uniform acceleration in \texttt{interIsoGoam}, implicit convection with \texttt{interfaceUpwind} gives errors of
0.02--0.03\,m/s,\footnote{\datanote{papers/part2/tab_fullSolver/fullSolver.txt},
case \texttt{implicitIU\_rhoNew}.} whereas explicit convection is exact
(Section~\ref{sec:timeLevel}, Tab.~\ref{tab:fullSolver}).
\item \emph{Uniform translation.} For a droplet translating with the ambient flow at a density
ratio of about $10^4$, they reported velocity errors $10^{10}$ to $10^{11}$ times larger than with
upwind for \texttt{limitedLinearV} and most other schemes, though not for QUICK (their Section~4.3
and Fig.~10). We do not reproduce these. A uniform
flow is preserved to round-off by every face interpolation at density ratios of $10^3$ and
$10^4$,\footnote{\datanote{papers/part2/tab_fullSolver/fullSolver.txt}, cases
\texttt{noForce\_*} and \texttt{noForceRatio1e4\_*}, with the largest deviation below $10^{-10}$\,m/s for upwind,
\texttt{limitedLinearV}, linear, QUICK and cubic, implicit and explicit.} as the analysis of
Section~\ref{sec:uniform} requires. Their tests with a velocity difference across the
interface, such as a droplet moving through air at rest (their Section~4.2), correspond to the mechanism of
Section~\ref{sec:faceValues}.
\end{itemize}

%% file: paper2/wave.tex
\section{A progressive wave}
\label{sec:wave}

We now test explicit convection and \texttt{interfaceUpwind} in the wave of Fig.~\ref{fig:problem}.
At larger time steps the wave gains energy,
which we partially counter with a Lax--Wendroff correction of explicit convection away from the interface
(Section~\ref{sec:LW}).

\parhead{Setup.} We use the progressive wave of Part~I \selfcite{RoenbyPart1}. It has a height of
0.125\,m, a period of 2\,s, a water depth of 0.4\,m and a wavelength of 3.79\,m, in a periodic
domain of one wavelength, with water and air at 1000 and 1\,kg/m$^3$. Two meshes are used, $190\times40$ and $380\times80$
cells (6 and 12 cells per wave height). The main runs use fixed time steps of 0.005\,s on the coarser and 0.0025\,s on the
finer mesh, which give a maximum Courant number of about 0.15. On the coarser mesh we also use
0.01\,s, to test a larger time step, and 0.0025\,s, to show how the momentum flux depends on the
time step. All runs use \texttt{interIsoGoam} with the geometric gravity model, the interface
reconstruction \texttt{isoAlpha} as in the wave of Part~I, Euler time stepping, one outer corrector, three pressure correctors and no momentum predictor. As in
Part~I, the runs are initialised with the exact wave of stream function theory, and they last
10\,s. We call this wave the exact input wave and compare the velocity with it. Its horizontal velocity at the surface of the crest,
0.445\,m/s, is the exact crest velocity.

\subsection{Crest velocity and spurious water current}
\label{sec:waveCrest}

\begin{figure}[tbp]
\centering
\includegraphics[width=0.79\linewidth]{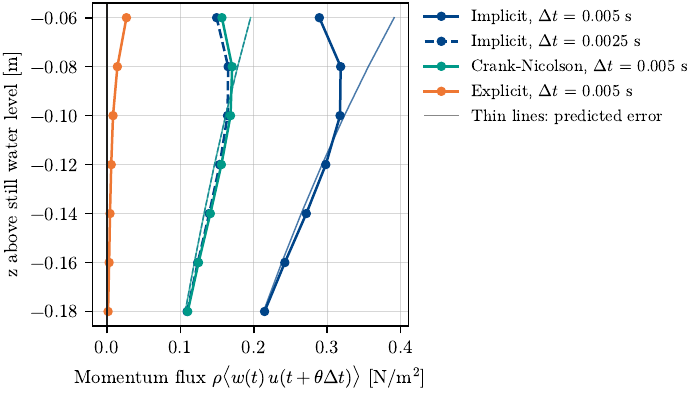}
\caption{Wave-averaged upward flux of horizontal momentum below the wave trough
($190\times40$ mesh, $2 < t \le 10$\,s), computed from the mass flux and the convected face velocity (symbols), and the lag prediction
(Eqn.~\ref{eq:lagFlux}) for the exact wave (thin lines). The exact flux
is zero. \figdir{fig_momFlux}.}
\label{fig:momFlux}
\end{figure}

\parhead{The spurious momentum flux.}
Below the trough, implicit convection produces an upward flux of horizontal momentum where
the exact wave has none (Fig.~\ref{fig:momFlux}). It agrees with the lag prediction
(Eqn.~\ref{eq:lagFlux}), evaluated with the exact wave, except close to the trough, and it halves
when the time step is halved. Crank--Nicolson halves it, and explicit convection removes it.

\begin{figure}[tbp]
\centering
\includegraphics[width=\linewidth]{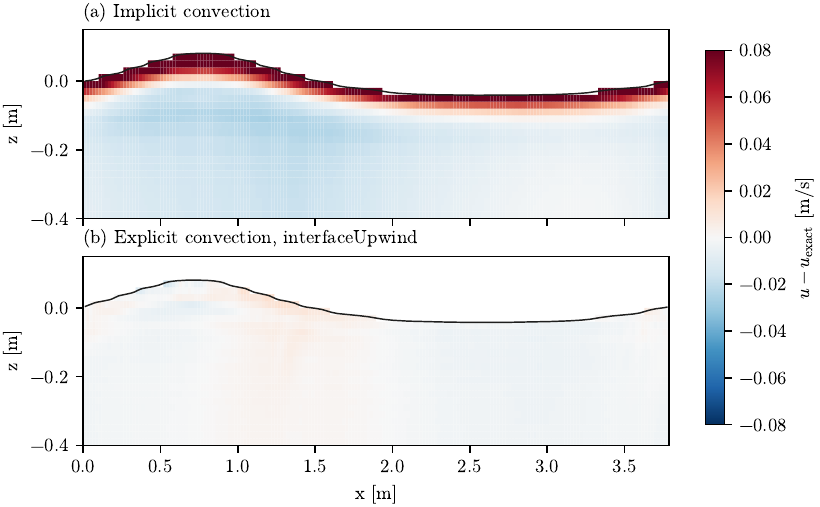}
\caption{Error of the horizontal velocity in the water at $t = 10$\,s ($190\times40$,
$\Delta t = 0.005$\,s), relative to the exact input wave at the crest position of the run. The air is not shown. The black line is the contour $\alpha = 0.5$.
\figdir{fig_waveField}.}
\label{fig:waveField}
\end{figure}

\begin{figure}[tbp]
\centering
\includegraphics[width=\linewidth]{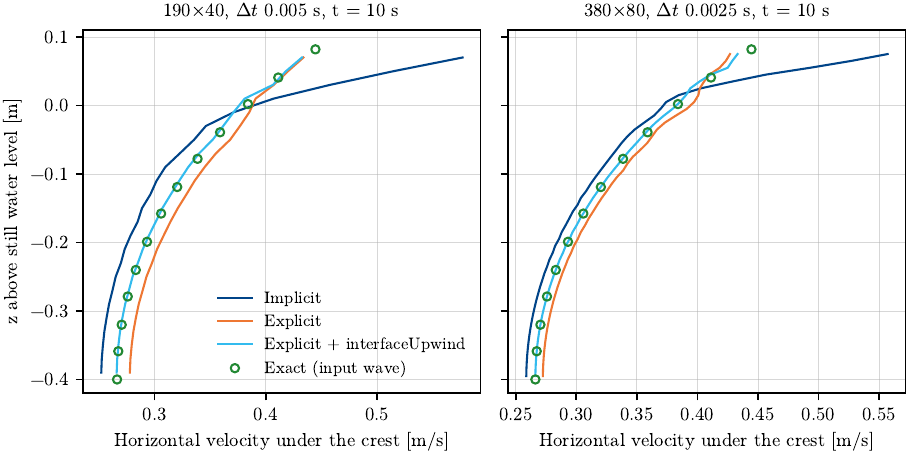}
\caption{Horizontal velocity under the crest at $t = 10$\,s and the exact input wave
(green circles). \figdir{fig_waveCrest}.}
\label{fig:waveCrest}
\end{figure}

\parhead{Crest velocity.}
With implicit convection the velocity under the crest is about 30\,\% too high after five
periods ($r_0$, Tab.~\ref{tab:wave}). Figure~\ref{fig:waveField} shows that the error is a layer
of excess velocity along the whole surface. This is a spurious water current near the surface, and below it
there is a weak return flow in the rest of the water column. Crank--Nicolson halves the error. Explicit convection removes it on
both meshes, and with \texttt{interfaceUpwind} the velocity in the whole crest column is within 2\,\%
of the exact value ($e_{\max}$ in Tab.~\ref{tab:wave}, and Figs.~\ref{fig:waveField}b and~\ref{fig:waveCrest}).

\begin{table}[tbp]
\centering
\caption{Progressive wave, $8 \le t \le 10$\,s. $\Delta H$ is the change of wave height [\%],
$r_0$ the horizontal velocity in the top full water cell under the crest relative to the exact
input wave, $e_{\max}$ the largest deviation of the velocity magnitude in the crest column [\% of
the exact crest velocity], $\bar u_s$ the wavelength-averaged velocity error in the top full water
cell, which is the surface current [m/s], and $\Delta c$ the celerity error [\%] relative to the input wave. The exact solution has
$\Delta H = 0$, $r_0 = 1$ and $e_{\max} = \bar u_s = \Delta c = 0$.
\figdir{tab_wave}.}
\label{tab:wave}
\small
\begin{tabular}{@{}l rccrr rccrr@{}}
\toprule
 & \multicolumn{5}{c}{$190\times40$, $\Delta t = 0.005$\,s} & \multicolumn{5}{c}{$380\times80$, $\Delta t = 0.0025$\,s} \\
\cmidrule(lr){2-6}\cmidrule(l){7-11}
Convection & $\Delta H$ & $r_0$ & $e_{\max}$ & $\bar u_s$ & $\Delta c$ & $\Delta H$ & $r_0$ & $e_{\max}$ & $\bar u_s$ & $\Delta c$ \\
\midrule
implicit                        & $-0.9$  & 1.31 & 30  & $+0.075$ & $+0.07$ & $-0.2$ & 1.26 & 25  & $+0.072$ & $+0.02$ \\
Crank--Nicolson                 & $0.0$   & 1.16 & 16  & $+0.038$ & $-0.06$ & $+0.3$ & 1.13 & 13  & $+0.037$ & $-0.05$ \\
explicit                        & $+4.2$  & 1.01 & 3.5 & $-0.011$ & $-0.11$ & $+2.5$ & 0.98 & 2.7 & $-0.013$ & $-0.08$ \\
explicit + interfaceUpwind      & $+0.3$  & 0.99 & 1.7 & $0.000$  & $-0.21$ & $+0.5$ & 0.99 & 1.1 & $-0.001$ & $-0.15$ \\
\bottomrule
\end{tabular}
\end{table}

\subsection{Wave height and energy}
\label{sec:waveEnergy}

Explicit convection with \texttt{limitedLinearV} lets the wave grow
($\Delta H$, Tab.~\ref{tab:wave}). \texttt{interfaceUpwind} removes most of the growth, but at larger
time steps the wave still grows.

\subsection{Forward Euler in the bulk: the Lax--Wendroff correction}
\label{sec:LW}

With a central convected face velocity, the convection operator conserves kinetic energy
\cite{Verstappen2003,Veldman2019}. Explicit convection advances it with forward Euler, which
adds kinetic energy every step, at a rate proportional to $\Delta t$. The Lax--Wendroff convected face velocity \cite{LaxWendroff1960} cancels this to
leading order by advancing the face velocity half a time step along the flow,
\begin{equation}
  \U_f = \mathcal{I}_f(\U^n) - \lambda_f\,\frac{\Delta t}{2}\,\bigl(\mathbf{u}_f\cdot\nabla\U^n\bigr)_f ,
  \label{eq:LW}
\end{equation}
where $\mathbf{u}_f$ and $(\nabla\U^n)_f$ are face interpolations whose face-normal parts
are evaluated as $\phi_f/|\mathbf{S}_f|$ and the compact face-normal gradient, respectively. The factor
$\lambda_f$ is the fraction of linear weighting in the underlying scheme. It is the limiter of
\texttt{limitedLinearV}, as in flux-limited Lax--Wendroff schemes \cite{Sweby1984}, and zero on the
upwind faces of \texttt{interfaceUpwind}. The correction therefore acts only away from the interface.
We denote it by LW. Explicit convection gives up the unconditional stability of implicit
convection (Section~\ref{sec:introduction}). For linear advection, forward Euler with a central
face value is unstable at every time step, whereas the Lax--Wendroff scheme is stable up to a
Courant number of one \cite{LaxWendroff1960,Ferziger2020}. None of the runs of
Section~\ref{sec:final}, at maximum Courant numbers up to 0.5, became unstable.

\begin{figure}[tbp]
\centering
\includegraphics[width=\linewidth]{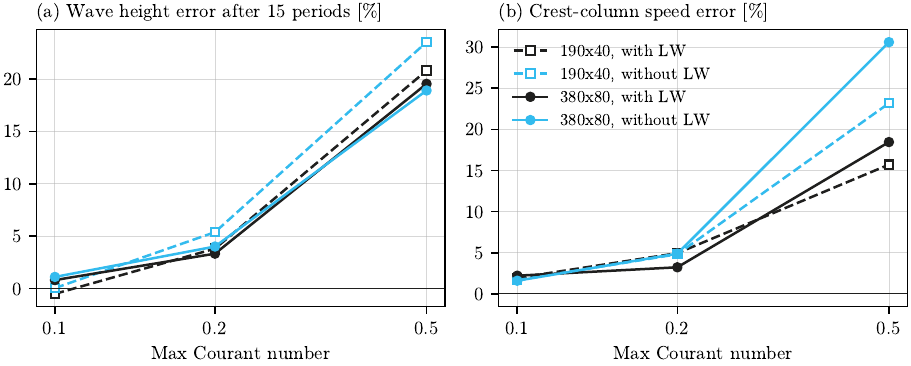}
\caption{Explicit convection with \texttt{interfaceUpwind}, with and without the Lax--Wendroff
correction: progressive wave over 15 periods (30\,s) under a Courant limit. (a) Wave height error after 15 periods. (b) Largest speed error in
the water column under the crest, relative to the exact crest velocity, mean over
28--30\,s. \figdir{fig_lwCheck}.}
\label{fig:lwCheck}
\end{figure}

Over 10\,s at the time steps used so far, the correction hardly changes the wave
height.\footnote{\datanote{papers/part2/tab_wave/wave.txt}.} Its effect shows over
longer times and at larger time steps. Over 15 periods under a Courant limit
(Fig.~\ref{fig:lwCheck}), it makes little difference at a maximum Courant number of 0.1.
At 0.2 the wave grows less with the LW correction. At 0.5 it grows by about 20\,\% with and
without the correction, but the velocity under the crest is clearly more accurate with it. The
correction therefore widens the range of Courant numbers in which the wave is accurate, but
it does not remove the growth at large Courant numbers. The origin of that growth is not
identified, and it limits the maximum Courant number to about 0.1 (Section~\ref{sec:final}).

The correction does not bring back the time lag of Section~\ref{sec:lag} in the wave. In the
translating Taylor--Green vortex it reintroduces half of the lag, because there
$\partial u/\partial t$ is due to advection alone, which the correction
contains.\footnote{\datanote{papers/part2/data/lag.txt}, case
\texttt{explicitLW\_dt0.01}.} In the wave, $\partial u/\partial t$ is set mainly by the pressure
gradient, which the correction does not contain. With the correction the surface current is at most
0.003\,m/s in magnitude, against 0.075\,m/s with implicit convection.\footnote{\datanote{papers/part2/tab_wave/wave.txt},
rows \emph{explicit + LW + interfaceUpwind}.} In the prescribed-flow tests, the correction neither reduces the errors
of explicit convection with \texttt{interfaceUpwind} nor spoils the exact uniform
acceleration.\footnote{\datanote{papers/part2/data/rig.txt}, rows
\emph{explicit, LW + \texttt{interfaceUpwind}}, and
\datasource{papers/part2/tab_fullSolver/fullSolver.txt}, case
\texttt{explicitLWIU\_rhoNew}.}

We recommend explicit convection with LW and \texttt{interfaceUpwind}. Unless stated otherwise, the \texttt{fluxinator} runs in the rest of this paper use these settings.

\subsection{Relation to the findings of Qwist and Christensen}
\label{sec:qwist}

Qwist and Christensen \cite{Qwist2023} compared \texttt{interFoam}, \texttt{interIsoFoam}, their own
implementation of the GFM, and their Direct Surface Description (DSD),
in which the air phase is removed, for a progressive wave. Their observations and
suggested explanations can be compared with the present results as follows.
\begin{itemize}
\item \emph{The crest velocity of \texttt{interIsoFoam} is 20--30\,\% too high, also on fine
meshes} (their Section~4.2.3). We find the same error and identify its cause. The lag flux (Eqn.~\ref{eq:lagFlux}) is
proportional to the time step and carries momentum up to the surface, where it accumulates in
a layer a few cells thick (Fig.~\ref{fig:waveField}a). The velocity excess of this layer
therefore scales with the time step divided by the cell size, that is, with the Courant number
rather than with the mesh. At equal Courant number it is
nearly the same on both meshes, and it increases when the time step is
doubled on the same mesh.\footnote{\datanote{papers/part2/tab_wave/wave.txt},
\emph{implicit}.} Their runs kept the Courant number constant under mesh refinement (their
Section~4.2), which explains why
the error did not decrease with refinement.
\item \emph{A sharp interface combined with a continuous velocity is inconsistent.} They
suggested that the sharp geometric advection of \texttt{interIsoFoam} is inconsistent with a
pressure and velocity that are continuous across the interface (their Section~4.2.2). As noted in Section~\ref{sec:introduction}, \texttt{interFoam}, whose interface is smeared, overpredicts
the crest velocity as well. Moreover, the present solver keeps the continuous velocity
of the one-fluid formulation, and with explicit convection and \texttt{interfaceUpwind} it reproduces the crest velocity to within 2\,\% ($e_{\max}$, Tab.~\ref{tab:wave}).
The inconsistencies are in the time level of the convected velocity and in the convected face velocity
at the interface, not in using one continuous velocity field for both fluids combined with a sharp interface.
\item \emph{The spurious velocities at a still interface may be the underlying problem} (their
Sections~1 and~4.1). Part~I removed the spurious air velocities in the tilted box without changing the
overprediction of the crest velocity. They are a separate error.
\item \emph{An accurate pressure is not sufficient.} Their GFM, which imposes the pressure
jump at the interface, also overpredicted the crest velocity (their Section~4.2.6). This agrees with the present
finding that the error lies in the convection term. Among their proposed candidate causes, the
convective discretisation, a linear-upwind scheme whose face values and gradients reach
across the interface, corresponds to the mechanism of Section~\ref{sec:faceValues}. The
temporal discretisation they used, the backward scheme, is inconsistent with the geometric
mass flux (Section~\ref{sec:uniform}). Their GFM with explicit fourth-order Runge--Kutta time
stepping, which has neither the backward scheme nor the time lag, also overpredicted the crest
velocity, which points to the convected face velocity at the interface. We have not run a GFM solver, so
for their GFM these are possible explanations, not verified ones.
\item \emph{The solver \texttt{interIsoFoam} failed at a smaller time step than \texttt{interFoam}} (their
Section~4.2.5). For \texttt{interIsoFoam} they used the backward scheme (their Section~4.2), which does not preserve even a uniform velocity across a moving interface
when combined with the geometric mass flux (Tab.~\ref{tab:rigUniform}). This is a likely
contributor, although we have not repeated their runs.
\item \emph{DSD reproduces the crest velocity closely.} The requirements of
Sections~\ref{sec:lag} and~\ref{sec:interface} are met in their method, as far as its description shows. Their explicit solvers
follow the Runge--Kutta method of Sanderse and Koren \cite{Sanderse2012}, in which each stage
evaluates the convection term with the velocity and the flux of that stage, so there is no time
lag (their Section~3.2.6). Their implicit DSD solver updates the convecting
flux in its outer iterations (their Section~3.2.3). The same holds for NS3 \cite{Mayer1998,Christensen2006}, which also solves only for the water and inspired DSD. Its Adams--Bashforth step pairs each face flux with the velocity of its own time level.
Without an air phase, the convected face velocities at the free surface come
from velocities extrapolated from the water, not from cells of the other phase. The present
results show that a similar accuracy can be obtained with the air phase retained, provided the
convection term meets these two requirements.
\end{itemize}

%% file: paper2/finalSolver.tex
\section{\texttt{fluxinator} on the wave study of Part I}
\label{sec:final}

\parhead{Setup.} \texttt{fluxinator} with the settings of Section~\ref{sec:LW} is run on the wave cases of Part~I \selfcite{RoenbyPart1} under a Courant limit. Four
meshes from $95\times20$ to $760\times160$ cells are used, each at maximum Courant numbers of 0.5, 0.2,
0.1 and 0.05, and every run lasts 30\,s (15 periods). Everything else is as in Part~I. The references are the
three solvers of Part~I. \texttt{interFoam} is run as released, with the \texttt{ddtCorr} term on, as
\OF users run it. \texttt{interIsoFoam} and \texttt{interIsoGoam} are rerun with the \texttt{ddtCorr} term
off (Part~I, Appendix~A), so that \texttt{interIsoGoam} and \texttt{fluxinator} differ only in the
momentum convection.\footnote{\datanote{papers/part2/fig_finalOverview/finalSolver.txt}. The runs are in
\datasource{run/progressiveWave}, studies \texttt{finalSolver\_maxCo}, \texttt{solvers\_maxCo}
(\texttt{interFoam}) and \texttt{solvers\_maxCo\_ddtCorrOff}.}

\noindent\texttt{fluxinator} (\datasource{applications/solvers/fluxinator}) is built from the same
source as \texttt{interIsoGoam} and differs only in its defaults: the geometric gravity model,
explicit convection, the \texttt{ddtCorr} term off and three reconstructions of the initial
interface before the first time step. With the same settings the two solvers therefore give
identical results. With the settings of Section~\ref{sec:LW}, \texttt{system/fvSchemes} of
\texttt{interIsoGoam} contains
\begingroup\small
\begin{verbatim}
divSchemes { div(rhoPhi,U) Gauss laxWendroff interfaceUpwind limitedLinearV 1; }
interfaceForces { gravity { model geometric; } }
\end{verbatim}
\endgroup
\noindent and \texttt{system/fvSolution} contains
\begingroup\small
\begin{verbatim}
PIMPLE { convectionNewTimeWeight 0; ddtCorr no; }
\end{verbatim}
\endgroup
\noindent The
entry \texttt{div(rhoPhi,U)} selects \texttt{limitedLinearV} with the treatment of the interface of
Section~\ref{sec:interface} and the correction of Section~\ref{sec:LW}, and
\texttt{convectionNewTimeWeight 0} gives explicit convection. In \texttt{fluxinator} only the
convection scheme has to be set, since the other entries are its defaults.\footnote{A case with
these settings: \datasource{run/progressiveWave/convection/explicitLWIU_190x40_dt0.005}.}

\begin{figure}[tbp]
\centering
\includegraphics[width=\linewidth]{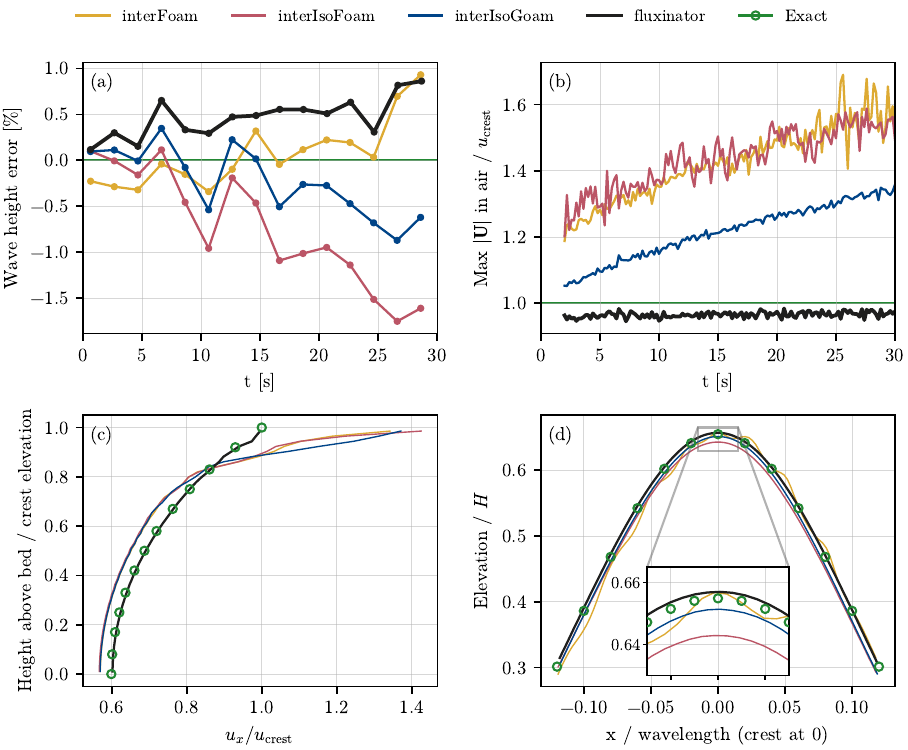}
\caption{Representative run: $380\times80$ mesh, maximum Courant number 0.1, \texttt{interFoam} as
released, \texttt{interIsoFoam} and \texttt{interIsoGoam} with the \texttt{ddtCorr} term off, and
\texttt{fluxinator}. (a) Wave height error of each wave. (b) Largest speed
in the air ($\alpha < 0.01$) at each written field, relative to the exact crest velocity. (c)
Horizontal velocity in the column nearest the crest at $t = 30$\,s, height relative to the exact
crest elevation. (d) Surface near the crest at $t = 30$\,s. The green circles are the exact input wave.
\figdir{fig_selCase}.}
\label{fig:selCase}
\end{figure}

\parhead{Representative run.} We show one run in detail, on the $380\times80$ mesh, which has about 12 cells per wave height,
at a maximum Courant number of 0.1 (Fig.~\ref{fig:selCase}).
On this mesh all four solvers keep the height of every wave to within 2\,\% over 15 periods, and
\texttt{fluxinator} maintains an accuracy similar to that of \texttt{interFoam} and \texttt{interIsoGoam}
(Fig.~\ref{fig:selCase}a). The difference is in the velocities. With \texttt{fluxinator} the largest speed in the air stays below the exact crest
velocity, while with the Part~I solvers it grows well above it (Fig.~\ref{fig:selCase}b).
Under the crest the Part~I solvers overshoot in the top cells, while \texttt{fluxinator} follows
the exact profile (Fig.~\ref{fig:selCase}c). The crests of all four solvers lie close to the exact
surface, but the surface of \texttt{interFoam} carries small undulations near the crest
(Fig.~\ref{fig:selCase}d), like the irregular crests of \texttt{interFoam} in Part~I.

\begin{figure}[tbp]
\centering
\includegraphics[width=\linewidth]{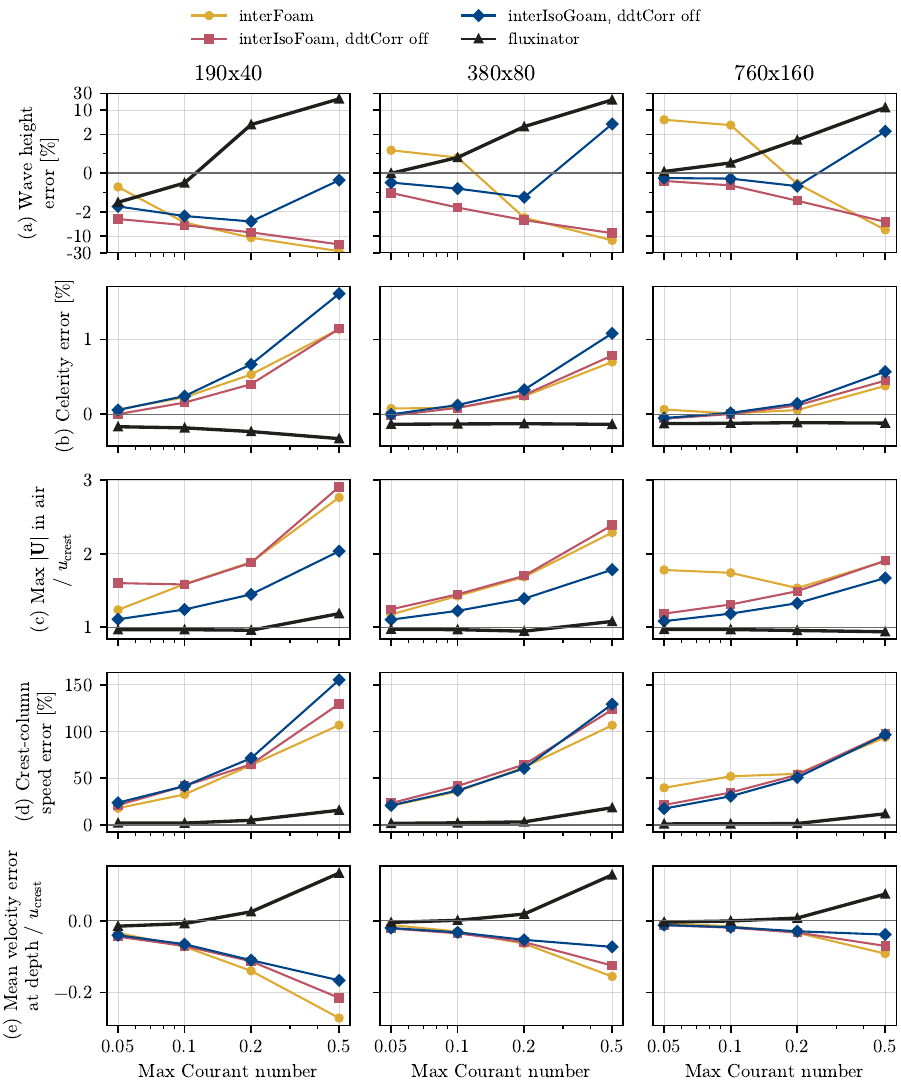}
\caption{Progressive wave of Part~I, 15 periods under a Courant limit: \texttt{fluxinator} (black),
\texttt{interFoam} as released, and \texttt{interIsoFoam} and \texttt{interIsoGoam} with the
\texttt{ddtCorr} term off. (a) Wave height error after 15 periods (non-linear scale, linear
between $-2$ and 2\,\%).
(b) Mean celerity error. (c) Median over $t > 2$\,s of the largest speed in the air
($\alpha < 0.01$), relative to the exact crest velocity. (d) Largest speed error in the
water column under the crest, relative to the exact crest velocity, mean over
28--30\,s. (e) Mean velocity error in the lower half of the same column (less than half the crest height above the bed),
relative to the exact crest velocity, 28--30\,s.
\figdir{fig_finalOverview}.}
\label{fig:finalOverview}
\end{figure}

\begin{figure}[tbp]
\centering
\includegraphics[width=\linewidth]{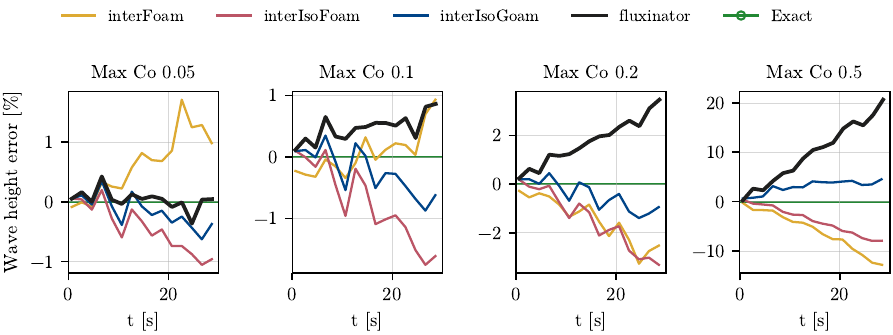}
\caption{Wave height error of each wave on the $380\times80$ mesh at the four maximum Courant
numbers. \figdir{fig_selHeight}.}
\label{fig:selHeight}
\end{figure}

\parhead{Wave height.}
At maximum Courant numbers of 0.1 and below, \texttt{fluxinator} keeps the wave height as
well as \texttt{interIsoGoam}, within about 1.5\,\% over 15 periods on the three finer meshes
(Fig.~\ref{fig:finalOverview}a). At 0.2, where \texttt{interIsoGoam} loses height, \texttt{fluxinator}
gains slightly, and at 0.5 the wave grows clearly. The error decreases steadily as the time
step is reduced.
The wave height of every wave on the $380\times80$ mesh (Fig.~\ref{fig:selHeight}) shows that
the growth at 0.2 and 0.5 builds up steadily from wave to wave, while at 0.1 and 0.05 \texttt{fluxinator} keeps the height at least as well as the Part~I solvers. \texttt{fluxinator} is therefore
meant for maximum Courant numbers of about 0.1 and below.

\parhead{Celerity.}
The Part~I solvers propagate the wave slightly too fast, more so at larger time steps
(Fig.~\ref{fig:finalOverview}b). \texttt{fluxinator} propagates it slightly too slowly, by
0.1--0.2\,\% at Courant numbers of 0.2 and below and nearly independently of the time step.
After 15 periods the crest therefore arrives about 0.04\,s too late. Since the error hardly
depends on the time step, it is an error of the spatial discretisation. It decreases slowly with
mesh refinement, from $-0.18$\,\% on the $190\times40$ mesh to $-0.13$ and $-0.12$\,\% on the
two finer meshes at a maximum Courant number of 0.1.

\parhead{Velocity in the air.}
With the Part~I solvers the largest speed in the air exceeds the exact crest velocity, by
up to a factor of three, and grows with the time step (Fig.~\ref{fig:finalOverview}c). With
\texttt{fluxinator} it stays below the crest velocity at Courant numbers of 0.2 and below, so the
air moves no faster than the water at the crest.

\parhead{Velocity in the water.}
Under the crest the speed errors of the Part~I solvers are 17--72\,\% of the exact
crest velocity at maximum Courant numbers of 0.2 and below, and they grow with the time step. Those of \texttt{fluxinator} are at most
5\,\% at Courant numbers of 0.2 and below, and at most 2\,\% at 0.1 and below
(Fig.~\ref{fig:finalOverview}d). The profiles of the
representative run (Fig.~\ref{fig:selCase}c) show the pattern. The Part~I solvers overshoot
in the top cells, and in the lower half of the water column they are slower than the exact wave
by a nearly uniform amount. This spurious mean water current against the direction of
propagation decreases as the mesh is refined and as the time step is reduced
(Fig.~\ref{fig:finalOverview}e). \texttt{fluxinator} follows the exact profile over the whole
depth at 0.2 and below.

\parhead{The coarsest mesh.}
The $95\times20$ mesh has three cells per wave height. On it, at maximum Courant numbers of 0.2 and below, \texttt{fluxinator}
loses 8--9\,\% of the wave height over 15 periods and has a speed error under the crest and a
current at depth, all nearly independent of the time step. Since the lag current is proportional to the
time step, these are errors of the spatial discretisation, which this paper does not address.
\texttt{interIsoFoam} and \texttt{interIsoGoam} also lose wave height on this mesh. The mesh is too coarse for this
wave.\footnote{\datanote{papers/part2/fig_finalOverview/finalSolver.txt},
rows $95\times20$.}

\begin{figure}[tbp]
\centering
\includegraphics[width=\linewidth]{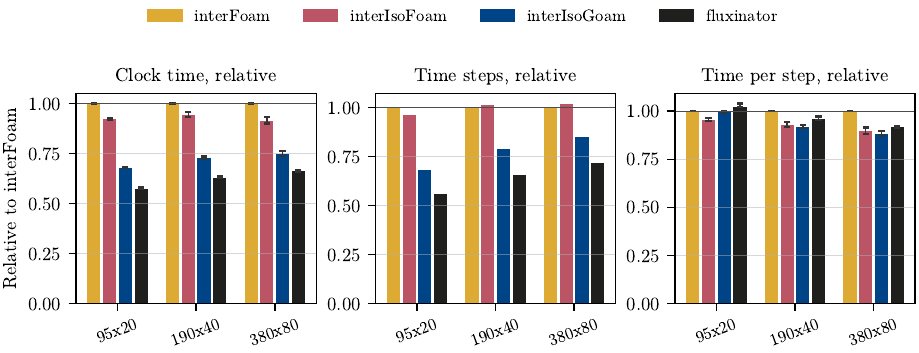}
\caption{Cost at a maximum Courant number of 0.1, relative to \texttt{interFoam} as released:
wall-clock time, number of time steps and time per step. The whiskers show the range over three
rounds of timing runs.
\figdir{fig_selCost}.}
\label{fig:selCost}
\end{figure}

\parhead{Cost.} Under a Courant limit, \texttt{fluxinator} needs fewer time steps than the
Part~I solvers, since the largest velocities, which are in the air and set the time step, are smaller.
At a maximum Courant number of 0.1 \texttt{fluxinator} needs 15--18\,\% fewer time steps than
\texttt{interIsoGoam}, and each step costs 2--6\,\% more. A step of \texttt{fluxinator} costs
4--8\,\% less than one of \texttt{interFoam} on the two finer meshes and 3\,\% more on the coarsest.
The wall-clock time of \texttt{fluxinator} is therefore 0.85--0.88 of that of \texttt{interIsoGoam}
and 0.58--0.66 of that of \texttt{interFoam} (Fig.~\ref{fig:selCost}).
We measured the time per step over $t = 2$--6\,s in serial runs, with the wave gauge switched off.
As in Part~I, the solvers compared ran at the same time on an otherwise idle machine, each process
bound to its own core and its memory allocated on the NUMA node of that core, with one run per NUMA node
and the clock frequency fixed. Each mesh was run in three such rounds, and the cost of a solver
is taken relative to \texttt{interFoam} of the same round. These ratios differ between the rounds
by at most 4\,\%. The wall-clock time is the number of time steps of the 30\,s run times the time
per step.\footnote{\datanote{run/progressiveWave/data/timing_maxCo0.1.csv}.}

%% file: paper2/conclusion.tex
\FloatBarrier
\section{Conclusions}
\label{sec:conclusion}

The overprediction of the crest velocity in water waves by the VoF solvers of \OF is caused by the momentum
convection term. Three requirements on the discretisation remove it.
\begin{itemize}
\item \emph{The convected velocity must belong to the time level of the mass flux.} The
interface advection computes the mass flux from the velocity at the start of the time step, but
the hardcoded implicit convection of \texttt{interFoam} and \texttt{interIsoFoam} convects the new velocity, from
the end of the step. In an unsteady flow, such as a travelling wave, this
lag produces a spurious flux of momentum proportional to the time step. In a progressive wave
it carries horizontal momentum upward, and a spurious current builds up in the water near the surface,
while a return flow develops below. The measured flux below the trough agrees with the lag prediction
(Section~\ref{sec:waveCrest}). The lag does not depend on the interface, and in a single-phase
Taylor--Green vortex moving with a uniform velocity it produces a spurious mean flow of the
predicted form. Explicit convection removes it.
\item \emph{The time derivative and the force must be consistent with the mass flux.} The
geometric mass flux of a time step satisfies the continuity equation in the two-level form of
an Euler step, so with this mass flux the time derivative must use the Euler scheme. The backward scheme does not preserve a uniform
velocity across a moving interface. A uniform acceleration is reproduced only if the force is
built with the density that belongs to the time level of the convected velocity. The gravity
force of \texttt{interIsoGoam}, like that of \texttt{interFoam} and \texttt{interIsoFoam}, uses the new density. It is consistent with explicit convection and
inconsistent with the implicit convection hardcoded in \texttt{interFoam} and \texttt{interIsoFoam}.
\item \emph{With explicit convection, the convected face velocity at the interface must be the upwind
value.} The new velocity of a cell is then a weighted average of old velocities, whatever the
density ratio, as long as a cell does not lose more mass in a step than it holds. A convected face velocity
that draws on the downwind cell amplifies velocity differences in cells that lose most of their
water in a step, by a factor that can reach the Courant number times the density ratio. In the
wave, upwind values at the interface also remove most of the growth of the wave height at small time steps. Upwind values are not exact, since the outflow leaves with the
mixed velocity of water and air in the cell.
\end{itemize}
The progressive wave of Section~\ref{sec:wave} was run on meshes with 6 and 12 cells per wave
height at a maximum Courant number of about 0.15. There, with explicit convection and \texttt{interfaceUpwind}, the horizontal velocity in the top water cell under the crest is within 1\,\% of the exact value after five periods, whereas implicit convection overpredicts it by
26--31\,\%. The surface current vanishes,
and the wave height changes by at most 0.5\,\% over the five periods. At larger time steps the
wave grows. A Lax--Wendroff correction away from the interface reduces this growth at a
maximum Courant number of 0.2 (Fig.~\ref{fig:lwCheck}) but does not remove it.

On the wave study of Part~I (Section~\ref{sec:final}), \texttt{fluxinator} reduces the speed
error under the crest from 17--72\,\% to at most 5\,\% at maximum Courant numbers of 0.2
and below, and to at most 2\,\% at 0.1 and below, keeps the air slower than the water at the crest, and removes the spurious water
current at depth. Its wave height is as accurate as that of \texttt{interIsoGoam} at maximum Courant
numbers of 0.1 and below, but grows at larger time steps. For the tested wave its celerity is
0.1--0.2\,\% too small and nearly independent of the time step. These statements hold on meshes with six or more cells
per wave height, whereas the coarsest mesh, with three, is too coarse for this wave. A time step
costs 2--6\,\% more than with implicit convection, but fewer steps are needed, since the spurious
velocities that limit the time step are removed. At a maximum Courant number of 0.1 the
wall-clock time over 15 wave periods is 0.85--0.88 of that of \texttt{interIsoGoam} and 0.58--0.66 of that of \texttt{interFoam}
as released.

\parhead{Recommendations.} For waves we recommend Euler time stepping with one outer
corrector, as in all runs of this paper, explicit convection (\texttt{convectionNewTimeWeight 0})
and the convection scheme
\begingroup\small
\begin{verbatim}
div(rhoPhi,U) Gauss laxWendroff interfaceUpwind limitedLinearV 1;
\end{verbatim}
\endgroup
\noindent At maximum Courant numbers of 0.1 and below the Lax--Wendroff correction makes
little difference, and \texttt{Gauss interfaceUpwind limitedLinearV 1} gives nearly the same
results. The \texttt{ddtCorr}
term should be off (\texttt{ddtCorr no}), the gravity force should use the geometric model of
Part~I, and the maximum Courant number should be about 0.1 or less for accurate wave
propagation. In \texttt{fluxinator} explicit convection, the \texttt{ddtCorr} term off and the geometric
gravity model are the defaults, so only the convection scheme and the time step have to be set.
These recommendations reflect the solver and the tests of this paper. The
solver and the methods are likely to be developed further, so users should check the code
repository and later publications for updated settings.

\parhead{Open questions.} At larger time steps a wave gain remains that increases steeply with the time step. Its origin is not
identified. Whether a second-order time scheme that is consistent with the mass flux, such as
the backward scheme with the flux combination of Section~\ref{sec:uniform}, reduces it has not
been tested. A convected face velocity at the interface that keeps the weighted average of upwind values
but is more accurate has not been found. The tests are two-dimensional and use uniform
meshes. The interface criterion of \texttt{interfaceUpwind} is formulated for general meshes but has
not been tested on them.
Finally, wave breaking is closely linked to the ratio of the horizontal particle velocity at
the crest to the celerity, and kinematic breaking criteria predict breaking when the particle
velocity reaches the celerity \cite{Perlin2013}. The solvers of the \texttt{interFoam} family are
known to break waves prematurely \cite{Jacobsen2012,Larsen2019}, and Larsen et
al.\ \cite{Larsen2019} suggested that the overpredicted crest velocity contributes to this.
Since the solver of this paper removes that overprediction while keeping the celerity nearly
exact, there is reason to expect that it also reduces premature breaking. This will be
investigated in future work.

%% file: parts/acknowledgements.tex
\section*{Acknowledgements}

J.R.\ was partially funded at Roskilde University by the InterFlow project, which is financed
by Independent Research Fund Denmark through a DFF Sapere Aude Research Leader grant (grant
number 9063-00018B).

%% file: parts/aiDeclaration.tex
\section*{Declaration of generative AI use}

\noindent The author used Claude Code with Anthropic's Claude Opus models extensively
in all aspects of this work, including the development of the solver and utilities, the
setup, running and post-processing of the test cases, the analysis of the results, the
preparation of figures and tables, the search for and checking of references, and the
drafting and editing of the text. All text, code and cases produced with the tool have
been thoroughly reviewed by the author, who takes full responsibility for the content of
this article, including all results and claims.